\documentclass[prd,preprintnumbers,nofootinbib]{revtex4}
\usepackage{graphicx}
\usepackage{color}
\usepackage{float}
\usepackage{color}
\usepackage{natbib}
\usepackage[plainpages=false, colorlinks=true, anchorcolor=blue, linkcolor=blue, citecolor=blue, bookmarks=false]{hyperref}
\newcommand{\rthis}[1]{\textcolor{black}{#1}}

\begin{document}
\title{Search for  Lorentz Invariance Violation using Bayesian model comparison applied to {\it Xiao et al.} GRB spectral lag catalog}
\author{Shantanu \surname{Desai}$^{1}$} \altaffiliation{E-mail: shntn05@gmail.com}

\author{Rajdeep \surname{Agrawal}$^{2}$}\altaffiliation{E-mail:ep18btech11012@iith.ac.in}
\author{Haveesh \surname{Singirikonda}$^{3}$}\altaffiliation{E-mail:haveesh.gss@gmail.com}

\affiliation{$^{1,2}$Department of Physics, Indian Institute of Technology, Hyderabad, Telangana-502285, India}
\affiliation{$^{3}$Argelander-Institut f\"ur Astronomie, Auf dem H\"ugel 71, 53121 Bonn, Germany }

\begin{abstract}
We use the  spectral lag  catalog of 46  short GRBs aggregated by Xiao et al~\cite{Xiao22}, to carry out an independent search for Lorentz Invariance violation (LIV).  For this purpose, we use a  power-law model as a function of energy for the intrinsic astrophysical  induced spectral lags. The expansion history of the universe  needed for constraining  LIV was obtained in a non-parametric method using cosmic chronometers.  We use Bayesian model comparison    to determine if the aforementioned spectral lags show evidence for LIV as compared to only astrophysically induced lags. We do not  find any evidence for LIV,  and obtain 95\% c.l. lower limits on the corresponding energy scale to be
$4 \times 10^{15}$ GeV and $6.8 \times 10^{9}$ GeV for the linear and quadratic LIV models respectively.  Our results obtained by using the  flat $\Lambda$CDM model for characterizing the cosmic expansion history are consistent with those obtained using chronometers.
\end{abstract}

\maketitle

\section{Introduction}
In some theoretical models beyond the Standard Model of Particle Physics, Lorentz Invariance is no longer an exact symmetry at energies close to the Planck scale ($E_{pl} \sim 10^{19}$ GeV), and
the speed of light varies as a function of energy~\cite{GAA98}: 
\begin{equation}
    v(E) = c\left[1 - s_{\pm}\frac{n+1}{2} \left(\frac{E}{E_{QG}}\right)^n\right],
    \label{eq:vE}
\end{equation}
where $s_{\pm} = \pm 1$ corresponds to sub-luminal ($s_{\pm}=+1$) or super-luminal ($s_{\pm}=-1$) Lorentz Invariance Violation (LIV). $E_{QG}$ denotes the energy scale where LIV effects dominate, and $n$ represents the order \rthis{of the modification of the photon group velocity}. In all LIV searches,  the series expansion is usually limited to linear ($n=1$) or quadratic corrections ($n=2$), because higher orders are impossible to reach experimentally.  Both linear and quadratic LIV models are predicted by different theoretical approaches,   which have been most recently reviewed   in~\citet{Addazi21}. 

For more than two decades Gamma-Ray Bursts (GRBs) have been a very powerful probe of LIV searches \cite{AmelinoCamelia98,Ellis03,Ellis,Abdo,Chang,Vasileiou13,Vasileiou15,Zhang,Liu,Pan15,Xu1,Xu2,Wei17,Ganguly,Ellis19,Weipolarization,Du,Pan,MAGIC,Zou,Agrawal21,Bartlett21,WeiWu2021,Liu22,Xiao22}. GRBs are single-shot explosions located at cosmological distances, which were first detected in 1960s  and have been observed over \rthis{ten} decades in energies from  keV to over 10~TeV range~\cite{Kumar}, \rthis{with the maximum GRB energy  equal to   18 TeV~\cite{LHAASO}}.  They are located at cosmological distances, although a distinct time-dilation signature in the light curves is yet to be seen~\cite{Singh}. GRBs are traditionally divided into two categories based on their durations, with  long (short) GRBs lasting more (less) than two seconds~\cite{Kouveliotou}.
Long GRBs are usually associated with core-collapse SN~\cite{Woosley} and short GRBs with neutron star mergers~\cite{Nakar}. There are however many exceptions to the aforementioned dichotomy, and many claims for additional GRB sub-classes have also been made~\cite{Kulkarni,Bhave} (and references therein).

The observable used in almost all the  LIV searches with GRBs consists of spectral lags,  defined as the arrival time difference  between high energy and low energy photons, and is positive if the high energy photons precede the low energy ones. A comprehensive review of all searches for LIV using GRB spectral lags (until 2021) can be found in our companion work~\cite{Agrawal21} (A21, hereafter). 

Most recently ~\citet{Xiao22} (X22, hereafter) carried out a search for LIV using a catalog of GRB spectral lags from {\it Swift} and {\it Fermi-GBM} detectors.
X22 constructed a dataset of spectral lags from 44 short GRBs  using {\it Swift} and 21 long GRBs using  {\it Fermi-GBM} data. The lags were obtained between the  fixed source frame energy intervals of 15-70 keV and 120-250 keV, and were obtained using  the novel Li-CCF method~\cite{Li04,Xiao21}.  This technique utilizes the temporal information in the light curve and is agnostic to the details of the cross-correlation function. The spectral lag data from these 46 short GRBs were used to look for LIV. A constant
source frame intrinsic lag was posited. Various criteria from information theory such as Akaike Information Criterion (AIC) and Bayesian Information Criterion (BIC)~\cite{Krishak4} were used to test for the significance of a putative  signature of LIV \rthis{and intrinsic emission}  compared to the null hypothesis of only an intrinsic spectral lag.  The intrinsic model as well as the LIV models provide equally good fit to the data. After assuming no evidence for LIV, a lower limit on $E_{QG}$ in the range $10^{15}-10^{17}$ GeV was obtained at 95\% c.l., depending on the value of $s_{\pm}$ and the model assumed for LIV.

In this work we improve upon the analysis in X22 in multiple ways:
\begin{itemize}
\item First, instead of a constant intrinsic spectral lag, we assume the same phenomenological model for the intrinsic spectral lag as in A21 (first used in Ref.~\cite{Wei}),   which was motivated from the results in ~\cite{Shao}. We note that the same type of phenomenological model for the intrinsic lags expressed as a power law of the energy has also been used in the  context of blazar flare modeling~\cite{Perennes20}. However, a lot more progress is also needed in source modeling to make a definitive case for the above power-law model for intrinsic emission. However, a constant intrinsic lag is unlikely, given the diversity among GRB light curves. Previous analyses which have used a constant intrinsic lag have had  incorporate an additional intrinsic scatter, which was kept as a free parameter or had to rescale the errors until the  reduced $\chi^2$ was close to one~\cite{Ellis,Wei17,Agrawal21}. A constant lag model is also nested within the power-law model, with a proper choice of parameters.

\item In the expression of LIV, instead of assuming the $\Lambda$CDM model, we use a non-parametric model-independent method for parameterizing the expansion history of the universe. The $\Lambda$CDM assumes the validity of General Relativity and Lorentz invariance. When searching for LIV, one should try to do this analysis in a model-independent way without resorting to an underlying theoretical model. Although the $\Lambda$CDM model agrees very well with the Planck CMB spectrum~\cite{Planck18}, in recent years several tensions with the $\Lambda$CDM model have  been found when considering low redshift data~\cite{Abdalla}. Therefore it makes sense to undertake this analysis for LIV   in using minimal theoretical assumptions.
However, for a comparison with our model-independent estimate, we also do a similar analysis using the $\Lambda$CDM model for the expansion history.
\item We incorporate the uncertainty due to the half-widths of both the energy intervals  in the likelihood used for the analysis. Since the light curves have been calculated using a finite energy interval, one should incorporate its half-width into the systematic error budget.
\item We use Bayesian model comparison for assessing the significance of LIV, instead of considerations based on information theory. Bayesian model comparison is more robust compared to AIC/BIC based tests~\cite{Trotta,Sanjib,Weller,Krishak4}. Furthermore, BIC is an approximation to the Bayesian evidence, after assuming a Gaussian distribution and in the limit of large sample size. 

\end{itemize}

The outline of this manuscript is as follows. We discuss the dataset and analysis methodology in Sect.~\ref{sec:data}. Our results are outlined in Sect.~\ref{sec:results} and we conclude in  Sect.~\ref{sec:conclusions}.

\section{Dataset and Analysis}
\label{sec:data}
\subsection{Data}
The dataset used for the analysis consists of spectral lags of 46 short GRBs collated  in X22 using the Li-CCF technique. \rthis{Note however that four of these ``short'' GRBs have T90 greater than two seconds, and hence based on the  classification criterion in ~\citet{Kouveliotou}, these objects should be considered as long GRBs. However, we should note that the  division at  two  seconds between short and long GRBs was based on BATSE data~\cite{Kouveliotou}. This dividing line  has been shown to be  detector-dependent and GRBs have also been shown to have  varying durations between the different detectors~\cite{Bromberg}. Among the four extra GRBs, three of them have T90 less than three seconds, whereas one of them has a duration close to eight seconds. These additional GRBs had redshift estimates and accurate estimates of spectral lag measurements and hence were included in our analysis similar to X22.}
The aforementioned sample spans the redshift range of $0.0098<z<2.609$.  The redshifts have been collated from multiple sources including the {\it Swift} Burst Analyzer~\cite{Evans09}, Fermi GBM Burst catalog~\cite{VonKienlin} as well as GCN circulars. The redshifts have been obtained using spectroscopy and their uncertainties are negligible, and hence not incorporated in our analyses. The spectral lags have been calculated using both {\it Swift} (44 GRBs) and {\it Fermi-GBM} (14 GRBs). X22 have shown that the spectral lags from both the detectors are consistent with each other within 1$\sigma$. For our analysis, we use the {\it Swift} derived spectral lags for 44 GRBs  and the {\it Fermi-GBM} lags for two GRBs, namely 200826A and 170817A. The spectral lags were computed as the difference between arrival times in the ranges 120-250 keV and 15-70 keV, with energies taken in the fixed rest frames of the sources.
To obtain the  uncertainties in the spectral lags, X22 carried out Monte Carlo simulations of the observed light curves, after positing a Gaussian and Poisson distribution for {\it Swift} and {\it Fermi-GBM}, respectively, and finally applying the Li-CCF technique. The $1\sigma$ uncertainties were obtained from the standard deviation of the resulting Gaussian distribution. More details on the error estimates  can be found in X22. In this work, these uncertainties are  used in the likelihood distribution need for calculating Bayesian evidence.

\subsection{Model for Spectral Lags}
The analysis methodology used in this work is the same as that used in A21.  
We briefly recap this procedure, while more details can be found in A21. The first step in our analysis involves constructing a model for the observed spectral lags.
This lag for a GRB located at redshift ($z$) can be written as the sum of two components:
\begin{equation}
\Delta t_{obs} = (1+z)\Delta t_{int}^{rest} + \Delta t_{LIV}
\label{eq:totaldelta}
\end{equation}
where
$\Delta t_{int}^{rest}$ is the intrinsic time delay due to astrophysical emission and $\Delta t_{LIV}$ is the lag due to LIV.
We use the following model for the intrinisc time lag~\cite{Wei}:
\begin{equation} 
\Delta t_{int}^{rest} =\tau\Big[ \Big(\frac{E}{keV}\Big)^{\alpha}-\Big(\frac{E_0}{keV}\Big)^{\alpha}\Big],
\label{eq:delt_int}
\end{equation}
where $E_0$ and $E$  correspond to the mid-point of the lower and upper energy intervals, namely $E_0 = 42.5$ keV and $E = 185$ keV.   This assumes that the energy spectrum is flat between these energy intervals. Actually, it's never the case, since GRB spectrum is usually described by the Band function~\cite{Band}. However, similar to A21 (and references therein), we also account for the finite energy interval  by incorporating them into the $1\sigma$ errors in $E_0$ and $E$. These errors in $E_0$ and $E$, corresponding to the half-widths of the upper and lower energy intervals, are equal to 27.5 keV and 65 keV, respectively. However, we should note that by assuming a flat spectrum, we are overestimating the errors. In a future work we shall also incorporate the GRB energy spectrum within the energy interval used for calculating the spectral lags.
In Eq.~\ref{eq:delt_int}, $\alpha$ represents the energy exponent and $\tau$ represents the time scale for the intrinsic time lag. Note that $\tau$ and $\alpha$ are considered as free parameters and evaluated later in this work.
This intrinsic model was empirically determined by modelling the single-pulse properties of about 50 GRBs~\cite{Shao} and has been widely used in a number of works~\cite{Wei,Wei2,Wei17,Ganguly,Pan,Du,Agrawal21}, and also  for blazar flares~\cite{Perennes20}. The constant spectral lag model used in X22 is nested within this power-law model. \rthis{We note that another assumption made is that the intrinsic lag is same for all GRBs.  Although, adding a GRB-dependent intrinsic parameters would be the most robust {\it ansatz}, this would increase the total number of free parameters leading to additional degeneracies. Since our sample mostly consists of short GRBs, assuming the same time lag would not be a totally egregious. However, given the observed  diversity in GRB light curves~\cite{Fishman95}, this assumption would definitely lead to some systematic effects, which are however hard to assess until we know the physics of the  intrinsic emission and its dependence on GRB properties. Alternatively, the spectral lags are also known to be correlated with GRB luminosities  and one could incorporate this lag-luminosity correlation while modelling the intrinsic time lags~\cite{Murase22}. We shall explore this in a future work.}

The LIV-induced lag is given by~\cite{Jacob}:
\begin{equation}
\Delta t_{LIV} =  -\left(\frac{1+n}{2H_0}\right)\left(\frac{E^n - E_0^n}{E_{QG,n}^n}\right)\frac{1}{(1+z)^n}\int_{0}^{z} \frac{(1+z^{\prime})^n}{h(z^{\prime})} \, dz^{\prime} 
\label{eq:deltaliv}
\end{equation} 
where $E_{QG,n}$ is the quantum gravity scale, above which LIV could have a significant effect; $E$ and $E_0$ have the  same  meaning as in Eq.~\ref{eq:delt_int}.   In Eq.~\ref{eq:deltaliv}, $n=1$ and  $n=2$ corresponds to  linear  and  quadratic LIV models, respectively.   This parametric form for $\Delta t_{LIV}$ corresponds to $s_{+}=1$ (cf. Eq.~\ref{eq:vE}). 
In Eq.~\ref{eq:deltaliv},
$h(z) \equiv \frac{H(z)}{H_0}$ is the dimensionless  Hubble parameter as a function of redshift. For the current standard $\Lambda$CDM model~\cite{Planck18}, $h(z)= \sqrt{\Omega_M (1+z^\prime)^3 + \Omega_\Lambda}$. This parametric form has been used in X22. In this work, we evaluate $h(z)$ using two methods. In the first method,
similar to A21, we have  evaluated the last term in the integrand non-parametrically using Gaussian Process Regression (GPR)~\cite{Seikel_2012}. The dataset used for GPR consists of cosmic chronometers. Cosmic chronometers provide a novel way to determine the Hubble parameter at different redshifts  using the relatives ages and spectroscopic redshifts of galaxies~\cite{Jimenez_2002}, and have been widely used in a number of cosmological analyses~\cite{Haveesh,BoraCDDR,Vagnozzi,Bora22} (and references therein). The only assumption involved in  this method  is that the universe is  described by the FLRW metric. The main systematic errors   which occur in the chronometer technique involve  errors in chronometer metallicity estimate, chronometer star formation history, uncertainty in stellar population synthesis models,   and rejuvenation effect. A detailed discussions of these systematic effects in the chronometer technique  as well as in sample selection can be found in the recent review~\cite{Moresco}.
We used the same chronometer dataset as in A21 (see also ~\cite{Haveesh}). Details of this non-parametric regression using GPR can be found in A21 and some of our other works ~\cite{Haveesh,BoraCDDR}. A comparison of GPR with other non-parametric regression techniques has also been carried out, and the results from other techniques have been found to be comparable to GPR within 5\%~\cite{Bora22}. 
In the second method we use  a flat $\Lambda$CDM  cosmology to evaluate $h(z)$ similar to X22. For this purpose, we assume $\Omega_m=0.3$, $\Omega_{\Lambda}=0.7$ and $H_0=70$ km/sec/Mpc.

Finally, we should note that the term in the integral in Eq.~\ref{eq:deltaliv} corresponds to one explicit model of LIV as evaluated in ~\citet{Jacob}, which assumes that the spacetime translations are not affected by quantum gravity scenarios. ~\citet{Rosati} have considered alternative LIV and doubly-special relativity models, where quantum gravity affects spacetime translations. Therefore, for this model, Eq.~\ref{eq:deltaliv} would no  longer be valid. Although, constraints on LIV for the \citet{Rosati} model have also been evaluated~\cite{Levy21}, in this work we shall only obtain limits on LIV using the Jacob-Piran model.

\subsection{Model Comparison}

We evaluate the significance of any LIV using Bayesian Model Comparison. To evaluate the significance of a model ($M_2$) as compared to another model ($M_1$), one usually calculates the Bayes factor ($B_{21}$) given by:
\begin{equation}
B_{21}=    \frac{\int P(D|M_2, \theta_2)P(\theta_2|M_2) \, d\theta_2}{\int P(D|M_1, \theta_1)P(\theta_1|M_1) \, d\theta_1} ,  \label{eq:BF}
\end{equation}
where $P(D|M_2,\theta_2)$ is the likelihood for the model $M_2$ given the data $D$ and $P(\theta_2|M_2)$ denotes the prior on the parameter vector $\theta_2$ of the model $M_2$.
The denominator in Eq.~\ref{eq:BF} denotes the same for model $M_1$. If $B_{21}$ is greater than one, then model 2 is preferred over model 1 and vice-versa. The significance can be qualitatively assessed using the Jeffreys' scale~\cite{Trotta}. More details on Bayesian model comparison can be found in various reviews~\cite{Trotta,Weller,Sanjib} as well as some of our past works~\cite{Haveesh,Krishak1,Krishak4,Srinikitha} in addition to A21.

In the present paper, the model $M_1$ corresponds to the hypothesis, where the spectral lags are produced    only by intrinsic astrophysical emission, whereas $M_2$ corresponds to the lags being described by Eq~\ref{eq:totaldelta}, consisting of both intrinsic and LIV delays. To calculate the Bayes factor, we need a model for the likelihood ($\mathcal{L}$), which we define as:
\begin{equation}
     \mathcal{L}=\prod_{i=1}^N \frac{1}{\sigma_{tot} \sqrt{2\pi}} \exp \left\{-\frac{[\Delta t _i-f(\Delta E_i,\theta)]^2}{2\sigma_{tot}^2}\right\},
     \label{eq:likelihood}
  \end{equation}
where $N$ is the total number of GRBs (46); $\Delta t_i$ denotes the observed spectral lag data, and where  $\sigma_{tot}$ denotes the total uncertainty which is given by 
\begin{equation} 
\sigma_{tot}^2 = \sigma_t^2 + \Big(\frac{\partial f}{\partial E}\Big)^2\sigma_E^2 + \Big(\frac{\partial f}{\partial E_0}\Big)^2\sigma_{E0}^2 \label{eq:totalerror}.
\end{equation}
In this expression, $f$ corresponds to  the particular model  being tested, \rthis{which could either be the two LIV models or the null hypothesis of only astrophysical emission}; $\sigma_t$ is the uncertainty on the spectral lag;  $\sigma_{E0}$ and $\sigma_E$ correspond to the  half-width of the lower and upper energy intervals, which are equal to 27.5 keV and 65 keV, respectively. \rthis{As mentioned earlier, we assume that the uncertainties in $E$ and $E_0$ are uncorrelated. We use the same time lags as in X22.}
Note that this uncertainty is much larger than the energy resolution of {\it Swift} and {\it Fermi-GBM} detectors which is about 100 eV and less than 10\%, respectively. In all previous analyses involving spectral lags where a  finite energy interval was used, both $E$ and $E_0$ have usually being chosen as the mid-point of the energy interval~\cite{Du,Agrawal21}. In principle, one could estimate a spectrum weighted value for $E$ and $E_0$, together with its uncertainties. However,  we have also included the size of the energy intervals into the uncertainties in $E$ and $E_0$, which are incorporated  in the likelihood. So our analysis can be considered  conservative.  As remarked earlier, we do not incorporate the error in spectroscopic redshift, as they are negligible. Since we are doing a  non-parametric regression of $H(z)$, it is not trivial to propagate the uncertainties in $H(z)$ due to systematic errors in chronometers. The dominant source of error in Eq.~\ref{eq:totalerror} is  the uncertainty in the spectral lag.

The last ingredient we need to evaluate Eq.~\ref{eq:BF} are the priors for the three models. We have used uniform priors for $\alpha$ and $\tau$, and log-uniform priors on $E_{QG}$. The prior ranges for all these parameters can be found in Table~\ref{priortable}. 

\section{Results}
\label{sec:results}
\subsection{Results using cosmic chronometers}
\begin{figure*}
    \centering
    \includegraphics[width=14cm,height=14cm]{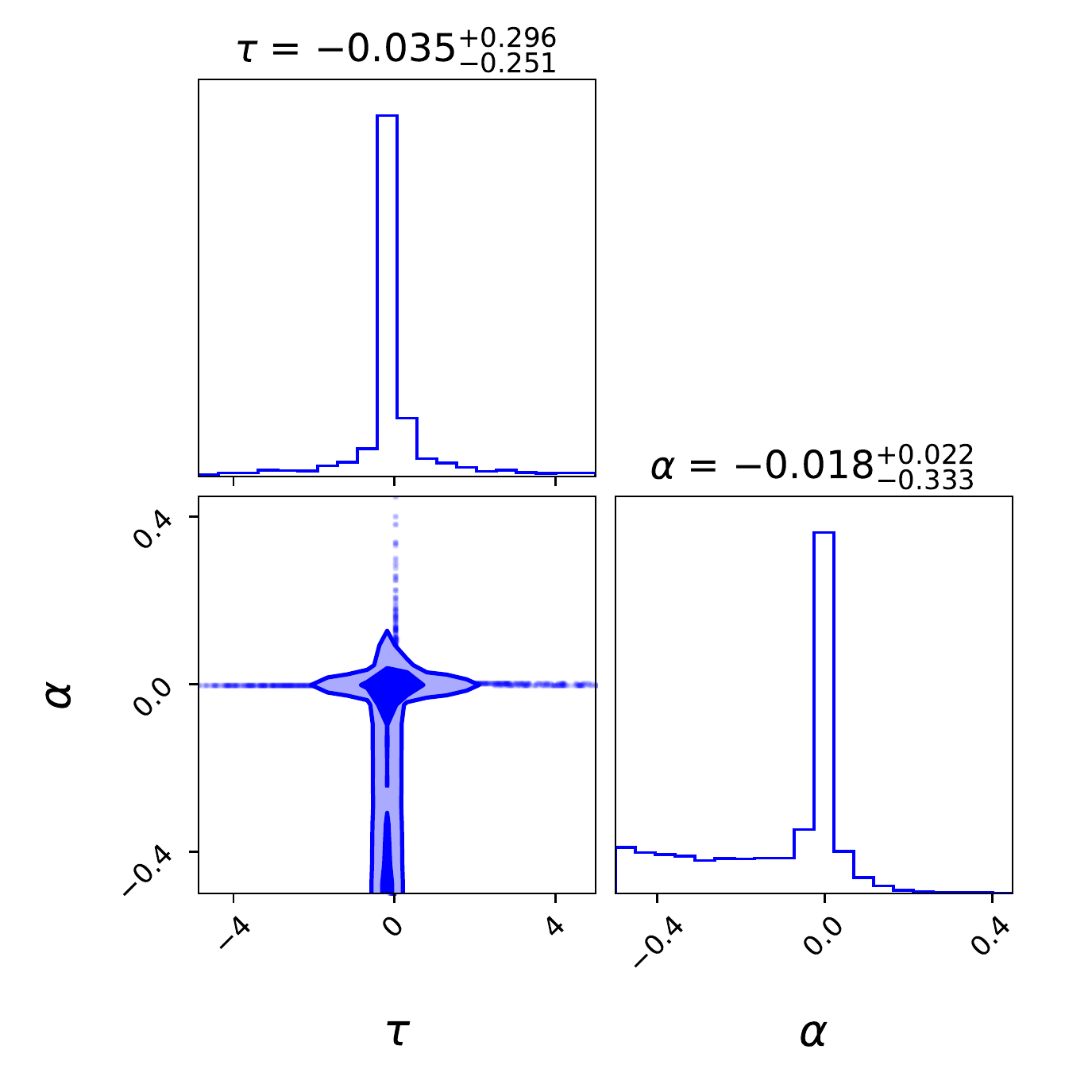}
    \caption{The marginalized 68\% and 90\% credible regions for the parameters of the null hypothesis of the observed time lags being only due to intrinsic emission.   The marginalized best-fit estimates for $\tau$ and $\alpha$ are depicted in the figure.}
    \label{fig:null}
\end{figure*}

\begin{figure*}
    \centering
    \includegraphics[width=15cm,height=15cm]{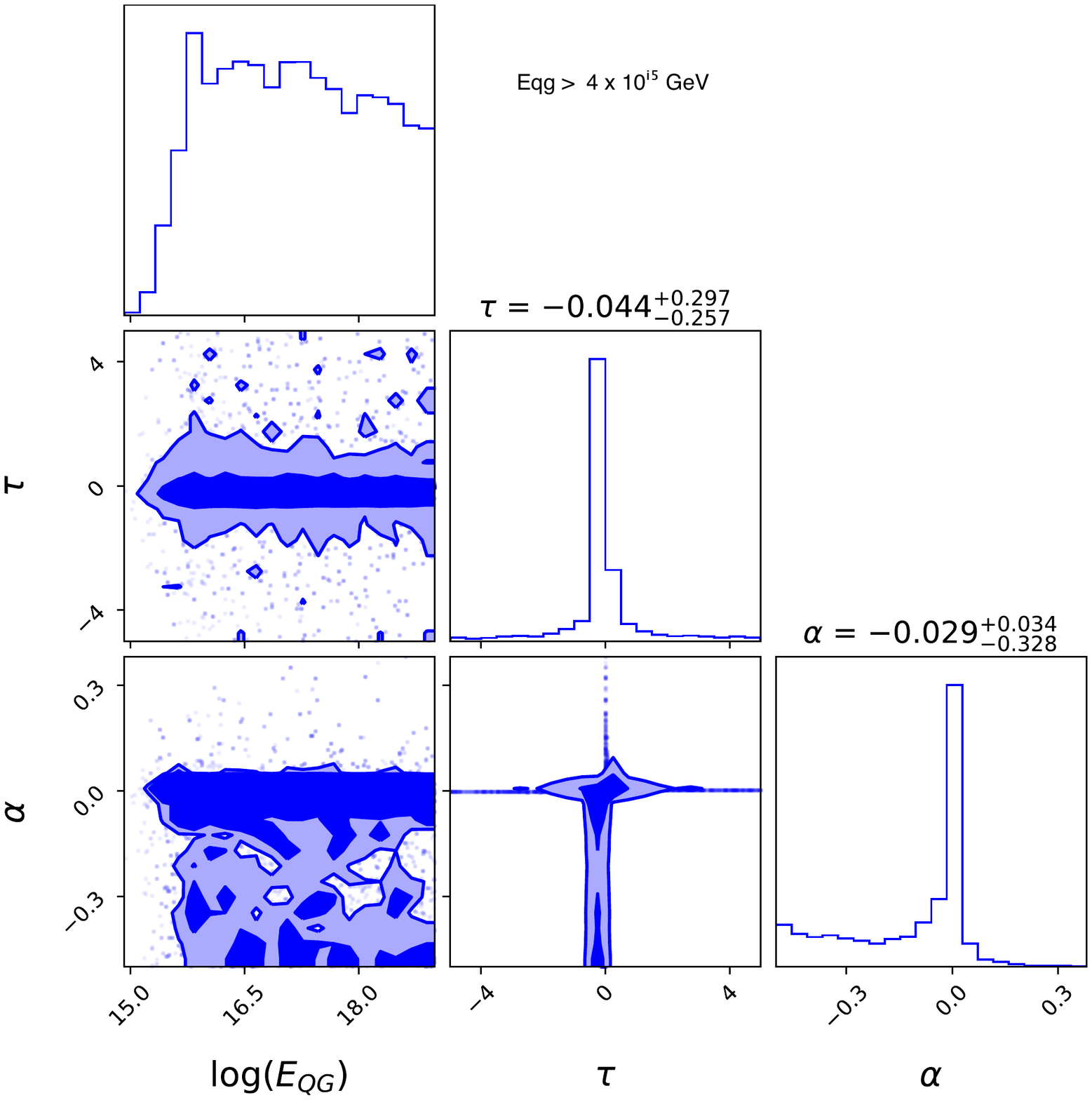}
    \caption{The marginalized 68\% and 90\% credible regions for the intrinsic lag along with the linear LIV model, corresponding to $n=1$   in Eq.~\ref{eq:deltaliv}. Since, no closed contour for $E_{QG}$ is obtained, we only set lower limits on $E_{QG}$ , given by $E_{QG}>4 \times 10^{15}$ GeV at 95\% c.l. For this figure, we use GPR to characterize the expansion history.}
    \label{fig:f1}
\end{figure*}

\begin{figure*}
    \centering
    \includegraphics[width=15cm,height=15cm]{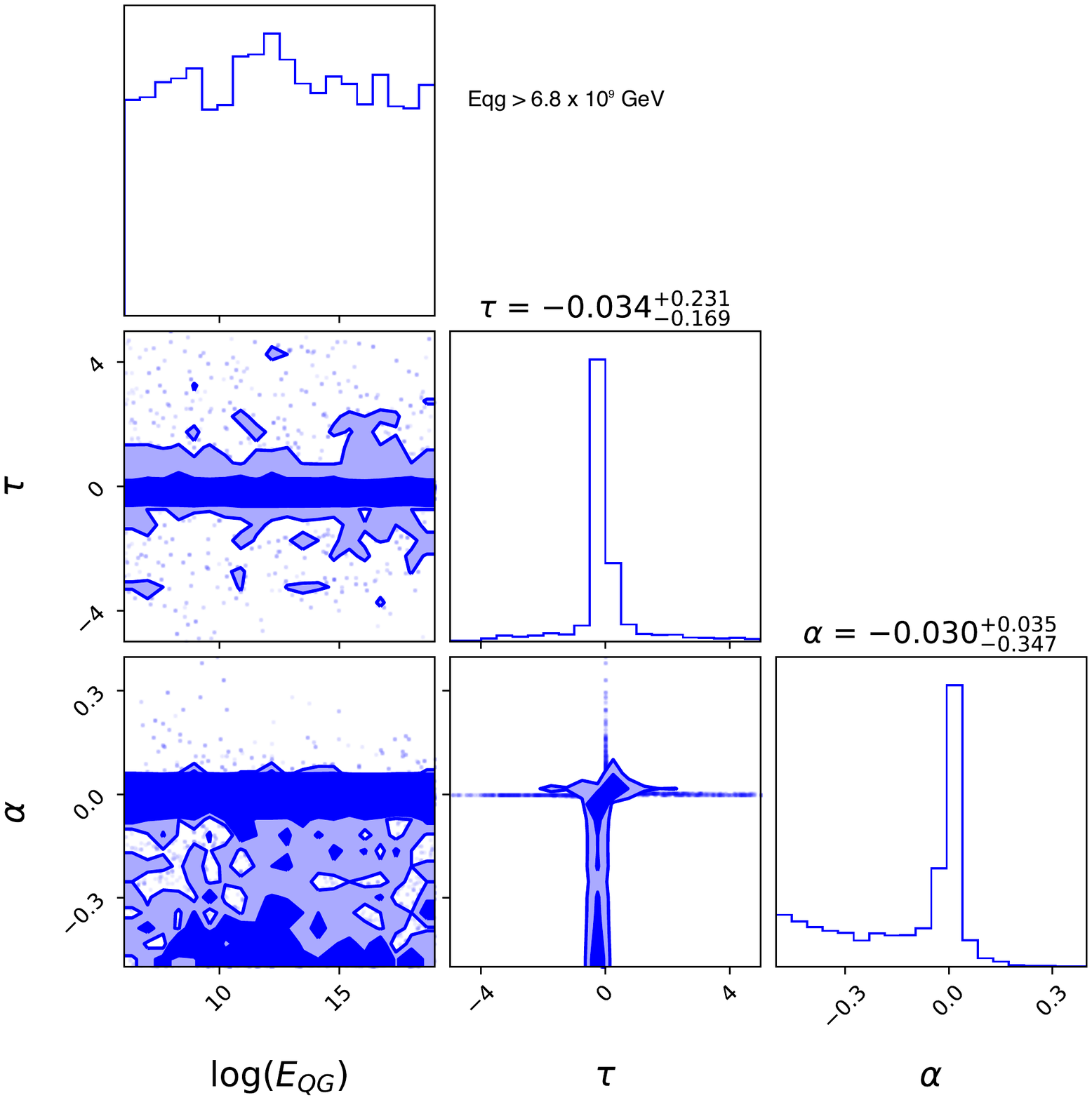}
    \caption{The marginalized 68\% and 90\% credible regions for the intrinsic lag along with the quadratic LIV model, corresponding to $n=2$   in Eq.~\ref{eq:deltaliv}. Since, no closed contour for $E_{QG}$ is obtained (similar to Fig.~\ref{fig:f1}), we only set lower limits on $E_{QG}$, given by $E_{QG}> 6.8 \times 10^9$ GeV at 95\% c.l. For this figure, we use GPR to characterize the expansion history. }
    \label{fig:f2}
\end{figure*}

Similar to A21, we used the Nested Sampling package {\tt dynesty}~\cite{dynesty} for calculating the Bayesian evidences for all the three models. The 68\% and 90\% marginalized credible intervals for the model parameters obtained from this sampling can be found  in Figs.~\ref{fig:null}, ~\ref{fig:f1}, ~\ref{fig:f2} for  the null hypothesis (consisting of only intrinsic lags), intrinsic lags along with  linear LIV model, and finally intrinsic lags along with quadratic LIV model, respectively. There is considerable degeneracy between $\alpha$ and $\tau$. We do not find closed contours for $\alpha$ and $\tau$ even at 2$\sigma$. The marginalized posterior for $\alpha$ is  also asymmetric and shows a long one side tail. The main reason  for this is due to the  large size of the uncertainties  in $\Delta$t.

The Bayes factor for the linear and quadratic LIV models along with the intrinsic lag  compared to the null hypothesis of only intrinsic emission,  as well as the $\chi^2/dof$ for all the three models can be found in Table~\ref{tab:results}. We   see that the Bayes factors for both the LIV + intrinsic emission hypotheses compared to the null hypothesis of only the intrinsic astrophysical emission  is close to one, indicating that there is no evidence for LIV. The $\chi^2/dof$ for all the three models is close to one, indicating that all of the models provide reasonably good fits.

We do not get closed contours for $E_{QG}$ for both the LIV models. Therefore, we set 95\%  c.l. lower limits using the same method as A21, following the prescription in ~\cite{Ellis,PDG}:
\begin{equation}
    \frac{\int \limits_{E_{QG}}^{E_{\infty}} L_{marg}(x) dx}{\int \limits_{E_{0}}^{E_{\infty}} L_{marg}(x) dx} =0.95, 
    \label{eq:lowerlimit}
\end{equation}
where  $L_{marg}(x)$ is the  likelihood  obtained after marginalizing over the nuisance parameters ($\tau$ and $\alpha$), and $E_{\infty}=10^{19}$ GeV, corresponding to the Planck scale. To evaluate Eq.~\ref{eq:lowerlimit}, we use the {\tt dynesty} package and the same priors as in Table~\ref{priortable}. The 95\% c.l. lower limit on $E_{QG}$  is  then given by   $E_{QG} > 4  \times 10^{15}$  GeV and $E_{QG} > 6.8 \times 10^{9}$  GeV for  linear  and quadratic LIV, respectively. 

\subsection{Results for $\Lambda$CDM}
\begin{figure*}
    \centering
    \includegraphics[width=15cm,height=15cm]{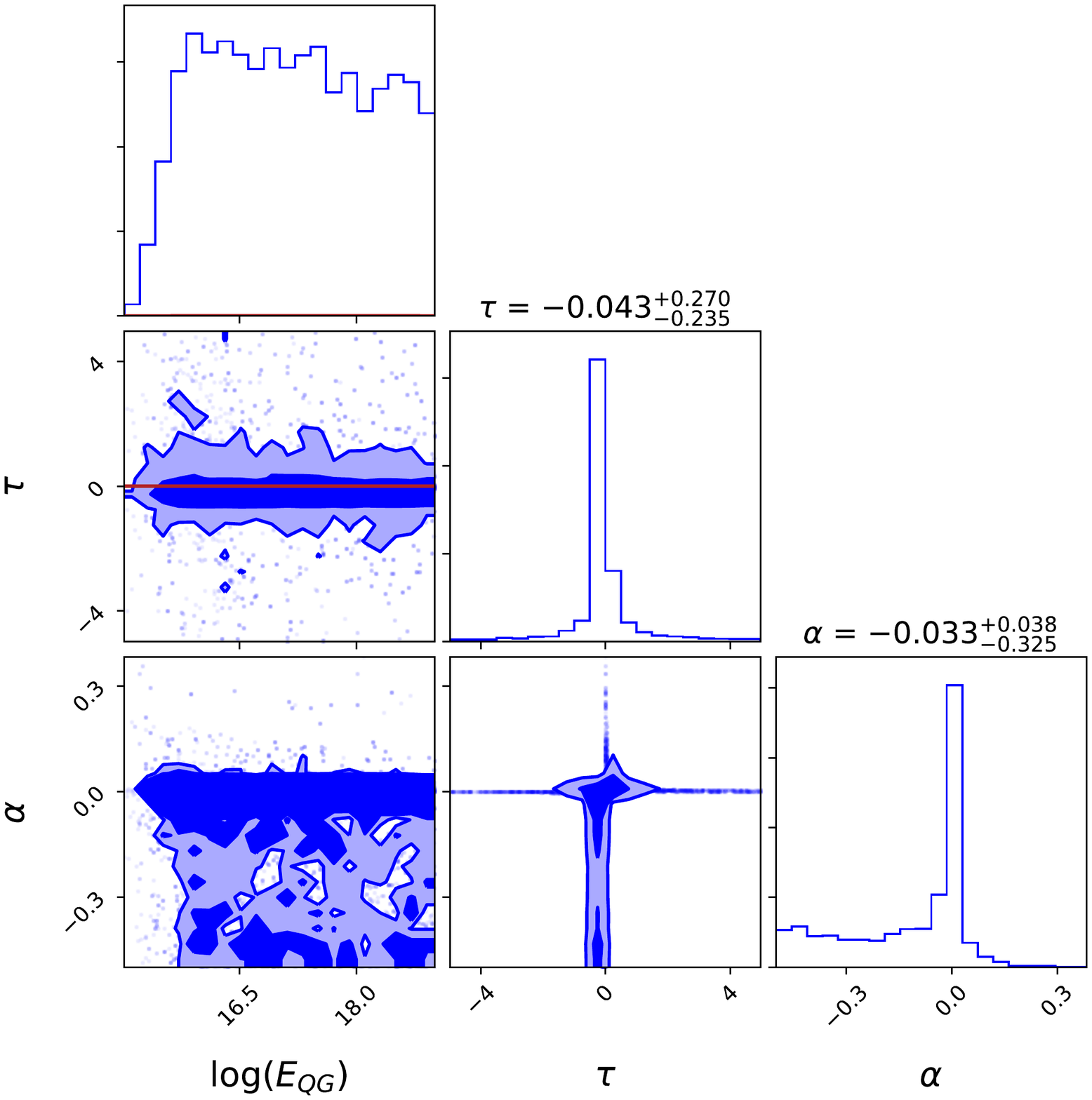}
    \caption{The marginalized 68\% and 90\% credible regions for the intrinsic lag along with the linear LIV model, corresponding to $n=1$   in Eq.~\ref{eq:deltaliv} and using the $\Lambda$CDM model to characterize the expansion history. The 95\% c.l. lower limit on $E_{QG}$ is given by $E_{QG}> 3.3 \times 10^{15}$ GeV.}
    \label{fig:f3}
\end{figure*}

\begin{figure*}
    \centering
    \includegraphics[width=15cm,height=15cm]{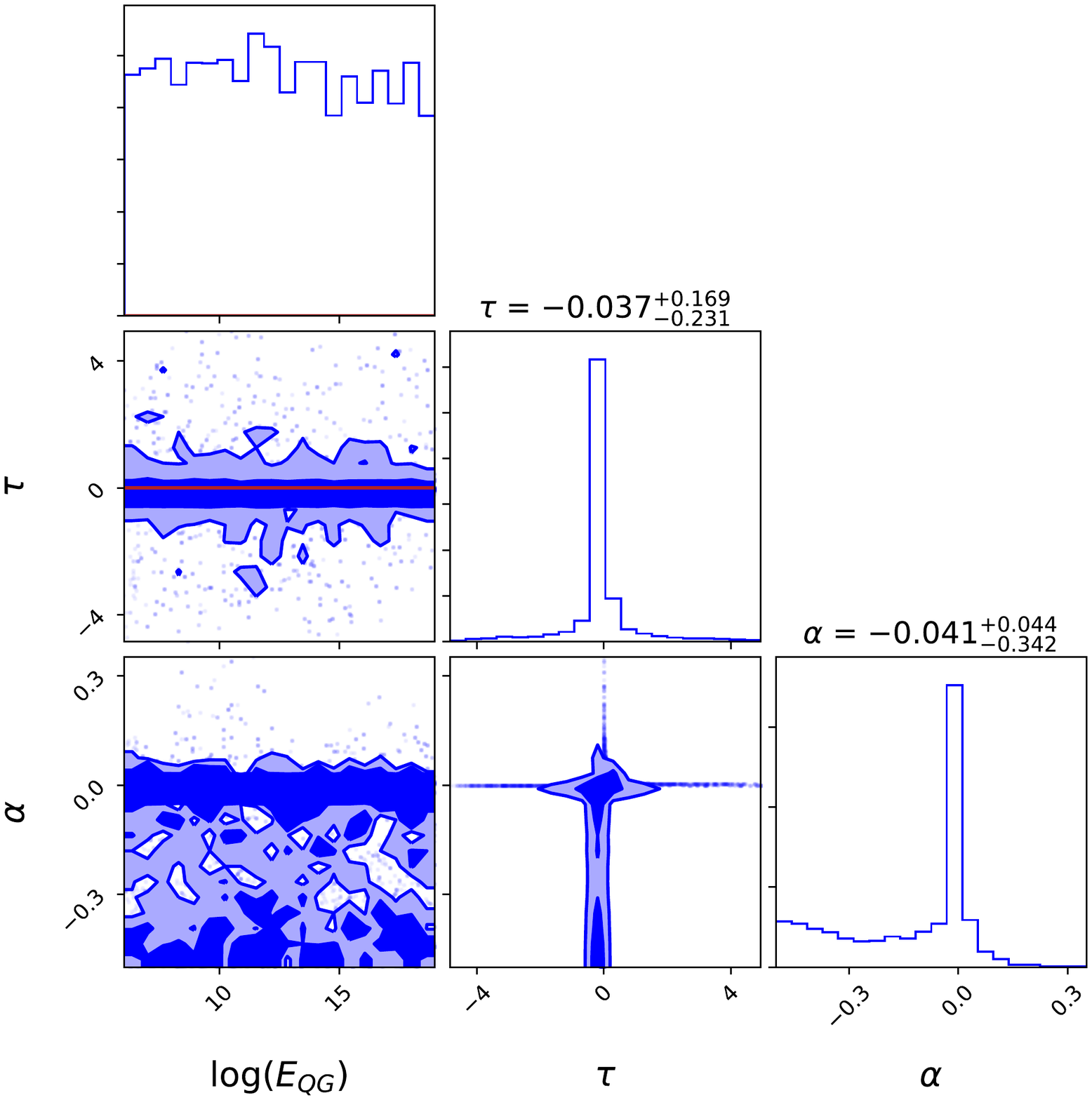}
    \caption{The marginalized 68\% and 90\% credible regions for the intrinsic lag along with the quadratic LIV model, corresponding to $n=2$   in Eq.~\ref{eq:deltaliv} and using  the $\Lambda$CDM model to characterize the expansion history. The 95\% c.l. lower limit on $E_{QG}$ is given by $E_{QG}>  10^{9}$ GeV.}
    \label{fig:f4}
\end{figure*}

As a crosscheck of our results hitherto obtained using a model-independent probe of expansion history, we redo our analyses by using the standard $\Lambda$CDM model to parameterize the expansion history
The corresponding results can be found in Figs.~\ref{fig:f3} and Figs~\ref{fig:f4}. Similar to the plots using chronometers, we do not get closed contours  for $E_{QG}$ for both the models.
The 95\% c.l. lower limit on $E_{QG}$  is  then given by   $E_{QG} > 3.3  \times 10^{15}$  GeV and $E_{QG} > 10^{9}$  GeV for  linear  and quadratic LIV, respectively. The Bayes factors for the linear and quadratic LIV models along with the intrinsic lag  compared to the null hypothesis of only intrinsic emission are close to one (cf. Table~\ref{tab:results}), indicating that there is no evidence for LIV. Once again the $\chi^2/dof$ are close to one for all the three models considered. Therefore, our results are consistent with those obtained using chronometers.

\begin{table}
    \centering
    
    \begin{tabular}{|c|c|c|c|}
    
    \hline
    \textbf{Parameter}&{\textbf{Prior}} & {\textbf{Minimum}} &  {\textbf{Maximum}}\\
        \hline 
    $\alpha$  & Uniform & -0.5 & 0.5  \\
    $\tau$ & Uniform & -5 & 5  \\
    $\log_{10} (E_{QG}/GeV)$  & Uniform & 6 & 19  \\
     \hline 
 \end{tabular}
 \caption{\label{priortable} Priors used for the calculation of Bayesian evidence for all the three models considered hitherto.}
\end{table}

\begin{table}
\begin{center}
\begin{tabular}{|c| c | c |  c | c|}
\hline
&  \textbf{No LIV} & \textbf{Expansion history} &  \textbf{ (n=1) LIV} & \textbf{(n=2) LIV}  \\
\hline
\(\chi^2/\rm{DOF} \) &  Chronometers & 55.8/44 & 51.3/43 & 52.6/43 \\

Bayes Factor  &   & -  & 0.3 &  1.1\\ \hline
\(\chi^2/\rm{DOF} \) &  $\Lambda$CDM & 55.8/44 & 50.6/43 & 50.2/43 \\
Bayes Factor  &   & -  & 0.3 &  1.03\\ \hline
\end{tabular}
\caption{\label{tab:results} Bayesian statistical significance of the two LIV models  as compared to the null hypothesis of only intrinsic emission. We also provide the $\chi^2$/DOF for all the three models, where DOF is equal to the total number of data points minus the number of free parameters. The first two rows show the results when cosmic chronometers were used to characterize the expansion history and the last two rows contain the results for $\Lambda$CDM. The Bayes factor shows negligible evidence for LIV for both the models and choices of expansion history. The $\chi^2$/DOF show reasonable fits for all the models.}
\end{center}
\end{table}

\subsection{Sensitivity to prior choices}
\rthis{ We should point out that for our analyses we have used uniform priors on all the three parameters. The prior  range for $\log(E_{QG}$ (GeV)) $\in [6,19]$ is a  conservative choice. Although some works have obtained limits on $E_{QG}<10^{19}$ GeV, these involve simplifying assumptions on the intrinsic time lags and hence we do not consider this.  Despite this conservative choice, we do not get closed contours for $E_{QG}$  in both the LIV analyses. We should also note that a few works have also obtained 95\% c.l. lower limits on $E_{QG}$ greater than the Planck scale~\cite{Abdo,Vasileiou13,Vasileiou15}. However, given that some works~\cite{Xu1,Xu2,Wei} (see also references in A21) have argued for evidence for LIV, contradicting the above lower limits, it is important to  test for signatures of LIV in order to verify these results.
We did check that choosing a Gaussian prior on $\tau$ with a mean equal to zero and $\sigma=0.3$ does not qualitatively change our conclusions.  
Therefore,  to summarize, although a detailed sensitivity  studies of our results as a function of prior choices is beyond the scope of this work, we have shown that our results do not change  with a Gaussian prior on $\tau$. Furthermore, since we not get closed contours for $E_{QG}$  despite choosing a wide prior range, they would not change our results if we truncate the prior.   }

\section{Conclusions}
\label{sec:conclusions}
In a recent work, X22 carried out a search for LIV from the  spectral lags of 46 short GRBs   between two fixed energy intervals in the source frame: 120-250 keV and 15-70 keV.

In this work, we carried out an independent search for LIV using the same spectral lag data following the methodology described in our previous work~\cite{Agrawal21}.  Instead of assuming a constant model for the intrinsic emission as in X22, we used the same power-law model as a function of energy  for the intrinsic emission as in A21. We parameterized the expansion history of the universe (needed to evaluate the LIV-induced lag) in a model-independent way using cosmic chronometers as well as using a flat $\Lambda$CDM cosmology. We searched for both linear and quadratic LIV.

The marginalized credible intervals for the parameters of all three of our models obtained using chronometers to characterize the expansion history  can be found in  Fig.~\ref{fig:null}, Fig.~\ref{fig:f1}, and Fig.~\ref{fig:f2}. 
Since we do not get closed contours for $E_{QG}$, we set 95\% c.l. lower limits. We find  that  $E_{QG} >4 \times 10^{15}$  GeV and $E_{QG} > 6.8 \times 10^{9}$ for  linear LIV and quadratic LIV, respectively. The results obtained using $\Lambda$CDM to calculate the  expansion history  are also comparable, viz  $E_{QG} >4 \times 10^{15}$  GeV and $E_{QG} > 6.8 \times 10^{9}$ for  linear LIV and quadratic LIV, respectively. The corresponding plots assuming  $\Lambda$CDM can be found in Fig.~\ref{fig:f3} and Fig.~\ref{fig:f4} and in Table~\ref{tab:results}.

 The corresponding 95\% c.l. lower limits obtained in X22 were  $\mathcal{O} (10^{15}-10^{17})$ GeV and 
$\mathcal{O} (10^6)$ GeV, for linear and quadratic LIV model respectively, depending on the sign  of the LIV term. We note however that X22 have obtained closed contours for the term which parameterizes the effect of LIV. Therefore, the one-sided lower limits reported in X22 do not adhere to the upper/lower limit estimation procedure recommended in PDG, and they should have instead reported the  bound confidence intervals for $E_{QG}$. There is a similar issue with many previous works in literature which have reported only one-sided lower limit on $E_{QG}$, despite obtaining bound confidence intervals  on  $E_{QG}$ which are less than the Planck scale\cite{Wei,Pan,Du}. For the stacked analysis in A21 using the datasets in ~\cite{Ellis,Wei,Du}, we did not get a closed contour for $E_{QG}$ for the linear LIV model (See Table 2 in A21). For this case the 68\% lower limit is given by $E_{QG}>10^{16}$ GeV. Therefore, our 95\% c.l. lower limit is comparable in magnitude to the corresponding result in A21.

The $\chi^2$/DOF for all the three models are close to one indicating that the current models  don't help to conclude on the presence of any of the effects investigated. We find that the Bayes factors are close to one, indicating that there is no evidence that spectral lags are induced due to LIV. 

We finally point out some limitations of our analysis. We have not considered the uncertainty in the cosmological  parameters or in the  chronometer $H(z)$ measurements. Although, previous works have shown that the results on LIV are insensitive to the underlying cosmological model~\cite{Pan15,Marek10}, it is still important to propagate the uncertainties due to Cosmology. Another limitation is that we have considered a flat spectrum for both the lower and upper energy intervals to calculate $E$ and $E_0$. Although this {\it ansatz} has been used in all works on LIV, for a more accurate estimate one must use the fitted Band spectrum parameters, to get the average value of $E$ and  $E_0$ along with its uncertainties. This will be implemented in a future work. Finally, the main
limiting source of systematic, which has been  the bane of all LIV searches is the uncertainty due to  the intrinsic GRB emission  mechanism~\cite{WeiWu2021}. The main assumption in this work is that the power-law model used for the intrinsic  emission mechanism (or the constant model used in X22 and other works) would adequately fit all GRBs in our sample with the same parameters. However, given the diversity in the light curves of GRBs, there is no guarantee that this assumption is correct. In the future one may need to do a separate analysis per GRB (based on the prompt or afterglow spectra) and apply phenomenological models such as that used in ~\cite{Chang12} \rthis{ or incorporate the luminosity-lag correlation~\cite{Murase22}} to model the intrinsic emission. Nevertheless, for these reasons it is more straightforward to carry out LIV analysis on a single GRB as done in ~\cite{Du,Wei}.

\rthis{We note that our limits are not as sensitive as the MAGIC limit  (from GRB 1900114C), which is  $\mathcal{O} (10^{19} GeV)$ and $\mathcal{O} (10^{10} GeV)$  for linear and quadratic LIV models, respectively~\cite{MAGIC} or the Fermi-LAT limit of $7.6 \times 10^{19}$/ $10^{11}$ GeV  for linear/quadratic LIV models obtained using four GRBs~\cite{Vasileiou13}. The sensitivity to LIV  increases  at higher photon energies. Since the MAGIC limit~\cite{MAGIC} is obtained from the detection of TeV gamma rays (as compared to keV energy GRBs in our work), their limit is the more stringent. However, this  result also involves  assumptions about the intrinsic spectral and temporal emission properties, for which conservative choices have been made in Ref.~\cite{MAGIC}. Similarly the limits in Ref.~\cite{Vasileiou13} were also obtained using GeV observations of GRBs, and hence are more stringent than those in our work. This work used three independent statistical methods. All these methods do involve assumptions regarding possible  source-intrinsic spectral-evolution effects, for which conservative choices have been made~\cite{Vasileiou13}. Therefore, to summarize, although the aforementioned works do not posit any parametric model for the intrinsic time delay (as in  our analysis), they do need to assume a model for the intrinsic spectral and temporal evolution in order to  obtain limits in $E_{QG}$.) }

We finally note that the broad Bayesian methodology used in this work to search for LIV can be easily applied to GRBs with spectral lag measurements in GeV energy range  as well as non-GRB sources, for which spectral lag measurements are available.

\section*{Acknowledgements}
\rthis{We are grateful to the anonymous referee for constructive feedback and comments on our manuscript.}
 
\bibliography{main}

\begin{thebibliography}{69}
\expandafter\ifx\csname natexlab\endcsname\relax\def\natexlab#1{#1}\fi
\expandafter\ifx\csname bibnamefont\endcsname\relax
  \def\bibnamefont#1{#1}\fi
\expandafter\ifx\csname bibfnamefont\endcsname\relax
  \def\bibfnamefont#1{#1}\fi
\expandafter\ifx\csname citenamefont\endcsname\relax
  \def\citenamefont#1{#1}\fi
\expandafter\ifx\csname url\endcsname\relax
  \def\url#1{\texttt{#1}}\fi
\expandafter\ifx\csname urlprefix\endcsname\relax\def\urlprefix{URL }\fi
\providecommand{\bibinfo}[2]{#2}
\providecommand{\eprint}[2][]{\url{#2}}

\bibitem[{\citenamefont{{Xiao} et~al.}(2022)\citenamefont{{Xiao}, {Xiong},
  {Wang}, {Zhang}, {Gao}, {Zhang}, {Cai}, {Yi}, {Zhao}, {Tuo} et~al.}}]{Xiao22}
\bibinfo{author}{\bibfnamefont{S.}~\bibnamefont{{Xiao}}},
  \bibinfo{author}{\bibfnamefont{S.-L.} \bibnamefont{{Xiong}}},
  \bibinfo{author}{\bibfnamefont{Y.}~\bibnamefont{{Wang}}},
  \bibinfo{author}{\bibfnamefont{S.-N.} \bibnamefont{{Zhang}}},
  \bibinfo{author}{\bibfnamefont{H.}~\bibnamefont{{Gao}}},
  \bibinfo{author}{\bibfnamefont{Z.}~\bibnamefont{{Zhang}}},
  \bibinfo{author}{\bibfnamefont{C.}~\bibnamefont{{Cai}}},
  \bibinfo{author}{\bibfnamefont{Q.-B.} \bibnamefont{{Yi}}},
  \bibinfo{author}{\bibfnamefont{Y.}~\bibnamefont{{Zhao}}},
  \bibinfo{author}{\bibfnamefont{Y.-L.} \bibnamefont{{Tuo}}},
  \bibnamefont{et~al.}, \bibinfo{journal}{\apjl}
  \textbf{\bibinfo{volume}{924}}, \bibinfo{eid}{L29} (\bibinfo{year}{2022}).

\bibitem[{\citenamefont{{Amelino-Camelia}
  et~al.}(1998{\natexlab{a}})\citenamefont{{Amelino-Camelia}, {Ellis},
  {Mavromatos}, {Nanopoulos}, and {Sarkar}}}]{GAA98}
\bibinfo{author}{\bibfnamefont{G.}~\bibnamefont{{Amelino-Camelia}}},
  \bibinfo{author}{\bibfnamefont{J.}~\bibnamefont{{Ellis}}},
  \bibinfo{author}{\bibfnamefont{N.~E.} \bibnamefont{{Mavromatos}}},
  \bibinfo{author}{\bibfnamefont{D.~V.} \bibnamefont{{Nanopoulos}}},
  \bibnamefont{and} \bibinfo{author}{\bibfnamefont{S.}~\bibnamefont{{Sarkar}}},
  \bibinfo{journal}{\nat} \textbf{\bibinfo{volume}{395}}, \bibinfo{pages}{525}
  (\bibinfo{year}{1998}{\natexlab{a}}).

\bibitem[{\citenamefont{Addazi et~al.}(2022)}]{Addazi21}
\bibinfo{author}{\bibfnamefont{A.}~\bibnamefont{Addazi}} \bibnamefont{et~al.},
  \bibinfo{journal}{Prog. Part. Nucl. Phys.} \textbf{\bibinfo{volume}{125}},
  \bibinfo{pages}{103948} (\bibinfo{year}{2022}), \eprint{2111.05659}.

\bibitem[{\citenamefont{{Amelino-Camelia}
  et~al.}(1998{\natexlab{b}})\citenamefont{{Amelino-Camelia}, {Ellis},
  {Mavromatos}, {Nanopoulos}, and {Sarkar}}}]{AmelinoCamelia98}
\bibinfo{author}{\bibfnamefont{G.}~\bibnamefont{{Amelino-Camelia}}},
  \bibinfo{author}{\bibfnamefont{J.}~\bibnamefont{{Ellis}}},
  \bibinfo{author}{\bibfnamefont{N.~E.} \bibnamefont{{Mavromatos}}},
  \bibinfo{author}{\bibfnamefont{D.~V.} \bibnamefont{{Nanopoulos}}},
  \bibnamefont{and} \bibinfo{author}{\bibfnamefont{S.}~\bibnamefont{{Sarkar}}},
  \bibinfo{journal}{\nat} \textbf{\bibinfo{volume}{393}}, \bibinfo{pages}{763}
  (\bibinfo{year}{1998}{\natexlab{b}}), \eprint{astro-ph/9712103}.

\bibitem[{\citenamefont{{Ellis} et~al.}(2003)\citenamefont{{Ellis},
  {Mavromatos}, {Nanopoulos}, and {Sakharov}}}]{Ellis03}
\bibinfo{author}{\bibfnamefont{J.}~\bibnamefont{{Ellis}}},
  \bibinfo{author}{\bibfnamefont{N.~E.} \bibnamefont{{Mavromatos}}},
  \bibinfo{author}{\bibfnamefont{D.~V.} \bibnamefont{{Nanopoulos}}},
  \bibnamefont{and} \bibinfo{author}{\bibfnamefont{A.~S.}
  \bibnamefont{{Sakharov}}}, \bibinfo{journal}{\aap}
  \textbf{\bibinfo{volume}{402}}, \bibinfo{pages}{409} (\bibinfo{year}{2003}),
  \eprint{astro-ph/0210124}.

\bibitem[{\citenamefont{{Ellis} et~al.}(2006)\citenamefont{{Ellis},
  {Mavromatos}, {Nanopoulos}, {Sakharov}, and {Sarkisyan}}}]{Ellis}
\bibinfo{author}{\bibfnamefont{J.}~\bibnamefont{{Ellis}}},
  \bibinfo{author}{\bibfnamefont{N.~E.} \bibnamefont{{Mavromatos}}},
  \bibinfo{author}{\bibfnamefont{D.~V.} \bibnamefont{{Nanopoulos}}},
  \bibinfo{author}{\bibfnamefont{A.~S.} \bibnamefont{{Sakharov}}},
  \bibnamefont{and} \bibinfo{author}{\bibfnamefont{E.~K.~G.}
  \bibnamefont{{Sarkisyan}}}, \bibinfo{journal}{Astroparticle Physics}
  \textbf{\bibinfo{volume}{25}}, \bibinfo{pages}{402} (\bibinfo{year}{2006}),
  \eprint{astro-ph/0510172}.

\bibitem[{\citenamefont{{Abdo} et~al.}(2009)\citenamefont{{Abdo}, {Ackermann},
  {Ajello}, {Asano}, {Atwood}, {Axelsson}, {Baldini}, {Ballet}, {Barbiellini},
  {Baring} et~al.}}]{Abdo}
\bibinfo{author}{\bibfnamefont{A.~A.} \bibnamefont{{Abdo}}},
  \bibinfo{author}{\bibfnamefont{M.}~\bibnamefont{{Ackermann}}},
  \bibinfo{author}{\bibfnamefont{M.}~\bibnamefont{{Ajello}}},
  \bibinfo{author}{\bibfnamefont{K.}~\bibnamefont{{Asano}}},
  \bibinfo{author}{\bibfnamefont{W.~B.} \bibnamefont{{Atwood}}},
  \bibinfo{author}{\bibfnamefont{M.}~\bibnamefont{{Axelsson}}},
  \bibinfo{author}{\bibfnamefont{L.}~\bibnamefont{{Baldini}}},
  \bibinfo{author}{\bibfnamefont{J.}~\bibnamefont{{Ballet}}},
  \bibinfo{author}{\bibfnamefont{G.}~\bibnamefont{{Barbiellini}}},
  \bibinfo{author}{\bibfnamefont{M.~G.} \bibnamefont{{Baring}}},
  \bibnamefont{et~al.}, \bibinfo{journal}{\nat} \textbf{\bibinfo{volume}{462}},
  \bibinfo{pages}{331} (\bibinfo{year}{2009}), \eprint{0908.1832}.

\bibitem[{\citenamefont{{Chang} et~al.}(2016)\citenamefont{{Chang}, {Li},
  {Lin}, {Sang}, {Wang}, and {Wang}}}]{Chang}
\bibinfo{author}{\bibfnamefont{Z.}~\bibnamefont{{Chang}}},
  \bibinfo{author}{\bibfnamefont{X.}~\bibnamefont{{Li}}},
  \bibinfo{author}{\bibfnamefont{H.-N.} \bibnamefont{{Lin}}},
  \bibinfo{author}{\bibfnamefont{Y.}~\bibnamefont{{Sang}}},
  \bibinfo{author}{\bibfnamefont{P.}~\bibnamefont{{Wang}}}, \bibnamefont{and}
  \bibinfo{author}{\bibfnamefont{S.}~\bibnamefont{{Wang}}},
  \bibinfo{journal}{Chinese Physics C} \textbf{\bibinfo{volume}{40}},
  \bibinfo{eid}{045102} (\bibinfo{year}{2016}), \eprint{1506.08495}.

\bibitem[{\citenamefont{{Vasileiou} et~al.}(2013)\citenamefont{{Vasileiou},
  {Jacholkowska}, {Piron}, {Bolmont}, {Couturier}, {Granot}, {Stecker},
  {Cohen-Tanugi}, and {Longo}}}]{Vasileiou13}
\bibinfo{author}{\bibfnamefont{V.}~\bibnamefont{{Vasileiou}}},
  \bibinfo{author}{\bibfnamefont{A.}~\bibnamefont{{Jacholkowska}}},
  \bibinfo{author}{\bibfnamefont{F.}~\bibnamefont{{Piron}}},
  \bibinfo{author}{\bibfnamefont{J.}~\bibnamefont{{Bolmont}}},
  \bibinfo{author}{\bibfnamefont{C.}~\bibnamefont{{Couturier}}},
  \bibinfo{author}{\bibfnamefont{J.}~\bibnamefont{{Granot}}},
  \bibinfo{author}{\bibfnamefont{F.~W.} \bibnamefont{{Stecker}}},
  \bibinfo{author}{\bibfnamefont{J.}~\bibnamefont{{Cohen-Tanugi}}},
  \bibnamefont{and} \bibinfo{author}{\bibfnamefont{F.}~\bibnamefont{{Longo}}},
  \bibinfo{journal}{\prd} \textbf{\bibinfo{volume}{87}}, \bibinfo{eid}{122001}
  (\bibinfo{year}{2013}), \eprint{1305.3463}.

\bibitem[{\citenamefont{{Vasileiou} et~al.}(2015)\citenamefont{{Vasileiou},
  {Granot}, {Piran}, and {Amelino-Camelia}}}]{Vasileiou15}
\bibinfo{author}{\bibfnamefont{V.}~\bibnamefont{{Vasileiou}}},
  \bibinfo{author}{\bibfnamefont{J.}~\bibnamefont{{Granot}}},
  \bibinfo{author}{\bibfnamefont{T.}~\bibnamefont{{Piran}}}, \bibnamefont{and}
  \bibinfo{author}{\bibfnamefont{G.}~\bibnamefont{{Amelino-Camelia}}},
  \bibinfo{journal}{Nature Physics} \textbf{\bibinfo{volume}{11}},
  \bibinfo{pages}{344} (\bibinfo{year}{2015}).

\bibitem[{\citenamefont{{Zhang} and {Ma}}(2015)}]{Zhang}
\bibinfo{author}{\bibfnamefont{S.}~\bibnamefont{{Zhang}}} \bibnamefont{and}
  \bibinfo{author}{\bibfnamefont{B.-Q.} \bibnamefont{{Ma}}},
  \bibinfo{journal}{Astroparticle Physics} \textbf{\bibinfo{volume}{61}},
  \bibinfo{pages}{108} (\bibinfo{year}{2015}), \eprint{1406.4568}.

\bibitem[{\citenamefont{{Liu} and {Ma}}(2018)}]{Liu}
\bibinfo{author}{\bibfnamefont{Y.}~\bibnamefont{{Liu}}} \bibnamefont{and}
  \bibinfo{author}{\bibfnamefont{B.-Q.} \bibnamefont{{Ma}}},
  \bibinfo{journal}{European Physical Journal C} \textbf{\bibinfo{volume}{78}},
  \bibinfo{eid}{825} (\bibinfo{year}{2018}), \eprint{1810.00636}.

\bibitem[{\citenamefont{{Pan} et~al.}(2015)\citenamefont{{Pan}, {Gong}, {Cao},
  {Gao}, and {Zhu}}}]{Pan15}
\bibinfo{author}{\bibfnamefont{Y.}~\bibnamefont{{Pan}}},
  \bibinfo{author}{\bibfnamefont{Y.}~\bibnamefont{{Gong}}},
  \bibinfo{author}{\bibfnamefont{S.}~\bibnamefont{{Cao}}},
  \bibinfo{author}{\bibfnamefont{H.}~\bibnamefont{{Gao}}}, \bibnamefont{and}
  \bibinfo{author}{\bibfnamefont{Z.-H.} \bibnamefont{{Zhu}}},
  \bibinfo{journal}{\apj} \textbf{\bibinfo{volume}{808}}, \bibinfo{eid}{78}
  (\bibinfo{year}{2015}), \eprint{1505.06563}.

\bibitem[{\citenamefont{{Xu} and {Ma}}(2016{\natexlab{a}})}]{Xu1}
\bibinfo{author}{\bibfnamefont{H.}~\bibnamefont{{Xu}}} \bibnamefont{and}
  \bibinfo{author}{\bibfnamefont{B.-Q.} \bibnamefont{{Ma}}},
  \bibinfo{journal}{Physics Letters B} \textbf{\bibinfo{volume}{760}},
  \bibinfo{pages}{602} (\bibinfo{year}{2016}{\natexlab{a}}),
  \eprint{1607.08043}.

\bibitem[{\citenamefont{{Xu} and {Ma}}(2016{\natexlab{b}})}]{Xu2}
\bibinfo{author}{\bibfnamefont{H.}~\bibnamefont{{Xu}}} \bibnamefont{and}
  \bibinfo{author}{\bibfnamefont{B.-Q.} \bibnamefont{{Ma}}},
  \bibinfo{journal}{Astroparticle Physics} \textbf{\bibinfo{volume}{82}},
  \bibinfo{pages}{72} (\bibinfo{year}{2016}{\natexlab{b}}),
  \eprint{1607.03203}.

\bibitem[{\citenamefont{{Wei} and {Wu}}(2017)}]{Wei17}
\bibinfo{author}{\bibfnamefont{J.-J.} \bibnamefont{{Wei}}} \bibnamefont{and}
  \bibinfo{author}{\bibfnamefont{X.-F.} \bibnamefont{{Wu}}},
  \bibinfo{journal}{\apj} \textbf{\bibinfo{volume}{851}}, \bibinfo{eid}{127}
  (\bibinfo{year}{2017}), \eprint{1711.09185}.

\bibitem[{\citenamefont{{Ganguly} and {Desai}}(2017)}]{Ganguly}
\bibinfo{author}{\bibfnamefont{S.}~\bibnamefont{{Ganguly}}} \bibnamefont{and}
  \bibinfo{author}{\bibfnamefont{S.}~\bibnamefont{{Desai}}},
  \bibinfo{journal}{Astroparticle Physics} \textbf{\bibinfo{volume}{94}},
  \bibinfo{pages}{17} (\bibinfo{year}{2017}), \eprint{1706.01202}.

\bibitem[{\citenamefont{{Ellis} et~al.}(2019)\citenamefont{{Ellis},
  {Konoplich}, {Mavromatos}, {Nguyen}, {Sakharov}, and
  {Sarkisyan-Grinbaum}}}]{Ellis19}
\bibinfo{author}{\bibfnamefont{J.}~\bibnamefont{{Ellis}}},
  \bibinfo{author}{\bibfnamefont{R.}~\bibnamefont{{Konoplich}}},
  \bibinfo{author}{\bibfnamefont{N.~E.} \bibnamefont{{Mavromatos}}},
  \bibinfo{author}{\bibfnamefont{L.}~\bibnamefont{{Nguyen}}},
  \bibinfo{author}{\bibfnamefont{A.~S.} \bibnamefont{{Sakharov}}},
  \bibnamefont{and} \bibinfo{author}{\bibfnamefont{E.~K.}
  \bibnamefont{{Sarkisyan-Grinbaum}}}, \bibinfo{journal}{\prd}
  \textbf{\bibinfo{volume}{99}}, \bibinfo{eid}{083009} (\bibinfo{year}{2019}),
  \eprint{1807.00189}.

\bibitem[{\citenamefont{{Wei}}(2019)}]{Weipolarization}
\bibinfo{author}{\bibfnamefont{J.-J.} \bibnamefont{{Wei}}},
  \bibinfo{journal}{\mnras} \textbf{\bibinfo{volume}{485}},
  \bibinfo{pages}{2401} (\bibinfo{year}{2019}), \eprint{1905.03413}.

\bibitem[{\citenamefont{{Du} et~al.}(2021)\citenamefont{{Du}, {Lan}, {Wei},
  {Zhou}, {Gao}, {Jiang}, {Zhang}, {Liu}, {Wu}, {Liang} et~al.}}]{Du}
\bibinfo{author}{\bibfnamefont{S.-S.} \bibnamefont{{Du}}},
  \bibinfo{author}{\bibfnamefont{L.}~\bibnamefont{{Lan}}},
  \bibinfo{author}{\bibfnamefont{J.-J.} \bibnamefont{{Wei}}},
  \bibinfo{author}{\bibfnamefont{Z.-M.} \bibnamefont{{Zhou}}},
  \bibinfo{author}{\bibfnamefont{H.}~\bibnamefont{{Gao}}},
  \bibinfo{author}{\bibfnamefont{L.-Y.} \bibnamefont{{Jiang}}},
  \bibinfo{author}{\bibfnamefont{B.-B.} \bibnamefont{{Zhang}}},
  \bibinfo{author}{\bibfnamefont{Z.-K.} \bibnamefont{{Liu}}},
  \bibinfo{author}{\bibfnamefont{X.-F.} \bibnamefont{{Wu}}},
  \bibinfo{author}{\bibfnamefont{E.-W.} \bibnamefont{{Liang}}},
  \bibnamefont{et~al.}, \bibinfo{journal}{\apj} \textbf{\bibinfo{volume}{906}},
  \bibinfo{eid}{8} (\bibinfo{year}{2021}), \eprint{2010.16029}.

\bibitem[{\citenamefont{{Pan} et~al.}(2020)\citenamefont{{Pan}, {Qi}, {Cao},
  {Liu}, {Liu}, {Geng}, {Lian}, and {Zhu}}}]{Pan}
\bibinfo{author}{\bibfnamefont{Y.}~\bibnamefont{{Pan}}},
  \bibinfo{author}{\bibfnamefont{J.}~\bibnamefont{{Qi}}},
  \bibinfo{author}{\bibfnamefont{S.}~\bibnamefont{{Cao}}},
  \bibinfo{author}{\bibfnamefont{T.}~\bibnamefont{{Liu}}},
  \bibinfo{author}{\bibfnamefont{Y.}~\bibnamefont{{Liu}}},
  \bibinfo{author}{\bibfnamefont{S.}~\bibnamefont{{Geng}}},
  \bibinfo{author}{\bibfnamefont{Y.}~\bibnamefont{{Lian}}}, \bibnamefont{and}
  \bibinfo{author}{\bibfnamefont{Z.-H.} \bibnamefont{{Zhu}}},
  \bibinfo{journal}{\apj} \textbf{\bibinfo{volume}{890}}, \bibinfo{eid}{169}
  (\bibinfo{year}{2020}), \eprint{2001.08451}.

\bibitem[{\citenamefont{Acciari et~al.}(2020)}]{MAGIC}
\bibinfo{author}{\bibfnamefont{V.~A.} \bibnamefont{Acciari}}
  \bibnamefont{et~al.} (\bibinfo{collaboration}{MAGIC, Armenian Consortium:
  ICRANet-Armenia at NAS RA, A. Alikhanyan National Laboratory, Finnish MAGIC
  Consortium: Finnish Centre of Astronomy with ESO}), \bibinfo{journal}{Phys.
  Rev. Lett.} \textbf{\bibinfo{volume}{125}}, \bibinfo{pages}{021301}
  (\bibinfo{year}{2020}), \eprint{2001.09728}.

\bibitem[{\citenamefont{{Zou} et~al.}(2018)\citenamefont{{Zou}, {Deng}, {Yin},
  and {Wei}}}]{Zou}
\bibinfo{author}{\bibfnamefont{X.-B.} \bibnamefont{{Zou}}},
  \bibinfo{author}{\bibfnamefont{H.-K.} \bibnamefont{{Deng}}},
  \bibinfo{author}{\bibfnamefont{Z.-Y.} \bibnamefont{{Yin}}}, \bibnamefont{and}
  \bibinfo{author}{\bibfnamefont{H.}~\bibnamefont{{Wei}}},
  \bibinfo{journal}{Physics Letters B} \textbf{\bibinfo{volume}{776}},
  \bibinfo{pages}{284} (\bibinfo{year}{2018}), \eprint{1707.06367}.

\bibitem[{\citenamefont{{Agrawal} et~al.}(2021)\citenamefont{{Agrawal},
  {Singirikonda}, and {Desai}}}]{Agrawal21}
\bibinfo{author}{\bibfnamefont{R.}~\bibnamefont{{Agrawal}}},
  \bibinfo{author}{\bibfnamefont{H.}~\bibnamefont{{Singirikonda}}},
  \bibnamefont{and} \bibinfo{author}{\bibfnamefont{S.}~\bibnamefont{{Desai}}},
  \bibinfo{journal}{\jcap} \textbf{\bibinfo{volume}{2021}}, \bibinfo{eid}{029}
  (\bibinfo{year}{2021}), \eprint{2102.11248}.

\bibitem[{\citenamefont{{Bartlett} et~al.}(2021)\citenamefont{{Bartlett},
  {Desmond}, {Ferreira}, and {Jasche}}}]{Bartlett21}
\bibinfo{author}{\bibfnamefont{D.~J.} \bibnamefont{{Bartlett}}},
  \bibinfo{author}{\bibfnamefont{H.}~\bibnamefont{{Desmond}}},
  \bibinfo{author}{\bibfnamefont{P.~G.} \bibnamefont{{Ferreira}}},
  \bibnamefont{and} \bibinfo{author}{\bibfnamefont{J.}~\bibnamefont{{Jasche}}},
  \bibinfo{journal}{\prd} \textbf{\bibinfo{volume}{104}}, \bibinfo{eid}{103516}
  (\bibinfo{year}{2021}), \eprint{2109.07850}.

\bibitem[{\citenamefont{{Wei} and {Wu}}(2021)}]{WeiWu2021}
\bibinfo{author}{\bibfnamefont{J.-J.} \bibnamefont{{Wei}}} \bibnamefont{and}
  \bibinfo{author}{\bibfnamefont{X.-F.} \bibnamefont{{Wu}}},
  \bibinfo{journal}{Frontiers of Physics} \textbf{\bibinfo{volume}{16}},
  \bibinfo{eid}{44300} (\bibinfo{year}{2021}), \eprint{2102.03724}.

\bibitem[{\citenamefont{{Liu} et~al.}(2022)\citenamefont{{Liu}, {Zhang}, and
  {Meng}}}]{Liu22}
\bibinfo{author}{\bibfnamefont{Z.-K.} \bibnamefont{{Liu}}},
  \bibinfo{author}{\bibfnamefont{B.-B.} \bibnamefont{{Zhang}}},
  \bibnamefont{and} \bibinfo{author}{\bibfnamefont{Y.-Z.}
  \bibnamefont{{Meng}}}, \bibinfo{journal}{arXiv e-prints}
  \bibinfo{eid}{arXiv:2202.09999} (\bibinfo{year}{2022}), \eprint{2202.09999}.

\bibitem[{\citenamefont{{Kumar} and {Zhang}}(2015)}]{Kumar}
\bibinfo{author}{\bibfnamefont{P.}~\bibnamefont{{Kumar}}} \bibnamefont{and}
  \bibinfo{author}{\bibfnamefont{B.}~\bibnamefont{{Zhang}}},
  \bibinfo{journal}{\physrep} \textbf{\bibinfo{volume}{561}},
  \bibinfo{pages}{1} (\bibinfo{year}{2015}), \eprint{1410.0679}.

\bibitem[{\citenamefont{{Huang} et~al.}(2022)\citenamefont{{Huang}, {Hu},
  {Chen}, {Zha}, {Liu}, {Yao}, {Cao}, and {Experiment}}}]{LHAASO}
\bibinfo{author}{\bibfnamefont{Y.}~\bibnamefont{{Huang}}},
  \bibinfo{author}{\bibfnamefont{S.}~\bibnamefont{{Hu}}},
  \bibinfo{author}{\bibfnamefont{S.}~\bibnamefont{{Chen}}},
  \bibinfo{author}{\bibfnamefont{M.}~\bibnamefont{{Zha}}},
  \bibinfo{author}{\bibfnamefont{C.}~\bibnamefont{{Liu}}},
  \bibinfo{author}{\bibfnamefont{Z.}~\bibnamefont{{Yao}}},
  \bibinfo{author}{\bibfnamefont{Z.}~\bibnamefont{{Cao}}}, \bibnamefont{and}
  \bibinfo{author}{\bibfnamefont{T.~L.} \bibnamefont{{Experiment}}},
  \bibinfo{journal}{GRB Coordinates Network} \textbf{\bibinfo{volume}{32677}},
  \bibinfo{pages}{1} (\bibinfo{year}{2022}).

\bibitem[{\citenamefont{{Singh} and {Desai}}(2022)}]{Singh}
\bibinfo{author}{\bibfnamefont{A.}~\bibnamefont{{Singh}}} \bibnamefont{and}
  \bibinfo{author}{\bibfnamefont{S.}~\bibnamefont{{Desai}}},
  \bibinfo{journal}{\jcap} \textbf{\bibinfo{volume}{2022}}, \bibinfo{eid}{010}
  (\bibinfo{year}{2022}), \eprint{2108.00395}.

\bibitem[{\citenamefont{{Kouveliotou} et~al.}(1993)\citenamefont{{Kouveliotou},
  {Meegan}, {Fishman}, {Bhat}, {Briggs}, {Koshut}, {Paciesas}, and
  {Pendleton}}}]{Kouveliotou}
\bibinfo{author}{\bibfnamefont{C.}~\bibnamefont{{Kouveliotou}}},
  \bibinfo{author}{\bibfnamefont{C.~A.} \bibnamefont{{Meegan}}},
  \bibinfo{author}{\bibfnamefont{G.~J.} \bibnamefont{{Fishman}}},
  \bibinfo{author}{\bibfnamefont{N.~P.} \bibnamefont{{Bhat}}},
  \bibinfo{author}{\bibfnamefont{M.~S.} \bibnamefont{{Briggs}}},
  \bibinfo{author}{\bibfnamefont{T.~M.} \bibnamefont{{Koshut}}},
  \bibinfo{author}{\bibfnamefont{W.~S.} \bibnamefont{{Paciesas}}},
  \bibnamefont{and} \bibinfo{author}{\bibfnamefont{G.~N.}
  \bibnamefont{{Pendleton}}}, \bibinfo{journal}{\apjl}
  \textbf{\bibinfo{volume}{413}}, \bibinfo{pages}{L101} (\bibinfo{year}{1993}).

\bibitem[{\citenamefont{{Woosley} and {Bloom}}(2006)}]{Woosley}
\bibinfo{author}{\bibfnamefont{S.~E.} \bibnamefont{{Woosley}}}
  \bibnamefont{and} \bibinfo{author}{\bibfnamefont{J.~S.}
  \bibnamefont{{Bloom}}}, \bibinfo{journal}{\araa}
  \textbf{\bibinfo{volume}{44}}, \bibinfo{pages}{507} (\bibinfo{year}{2006}),
  \eprint{astro-ph/0609142}.

\bibitem[{\citenamefont{{Nakar}}(2007)}]{Nakar}
\bibinfo{author}{\bibfnamefont{E.}~\bibnamefont{{Nakar}}},
  \bibinfo{journal}{\physrep} \textbf{\bibinfo{volume}{442}},
  \bibinfo{pages}{166} (\bibinfo{year}{2007}), \eprint{astro-ph/0701748}.

\bibitem[{\citenamefont{{Kulkarni} and {Desai}}(2017)}]{Kulkarni}
\bibinfo{author}{\bibfnamefont{S.}~\bibnamefont{{Kulkarni}}} \bibnamefont{and}
  \bibinfo{author}{\bibfnamefont{S.}~\bibnamefont{{Desai}}},
  \bibinfo{journal}{\apss} \textbf{\bibinfo{volume}{362}}, \bibinfo{eid}{70}
  (\bibinfo{year}{2017}), \eprint{1612.08235}.

\bibitem[{\citenamefont{{Bhave} et~al.}(2022)\citenamefont{{Bhave}, {Kulkarni},
  {Desai}, and {Srijith}}}]{Bhave}
\bibinfo{author}{\bibfnamefont{A.}~\bibnamefont{{Bhave}}},
  \bibinfo{author}{\bibfnamefont{S.}~\bibnamefont{{Kulkarni}}},
  \bibinfo{author}{\bibfnamefont{S.}~\bibnamefont{{Desai}}}, \bibnamefont{and}
  \bibinfo{author}{\bibfnamefont{P.~K.} \bibnamefont{{Srijith}}},
  \bibinfo{journal}{\apss} \textbf{\bibinfo{volume}{367}}, \bibinfo{eid}{39}
  (\bibinfo{year}{2022}), \eprint{1708.05668}.

\bibitem[{\citenamefont{{Li} et~al.}(2004)\citenamefont{{Li}, {Qu}, {Feng},
  {Song}, {Ding}, and {Chen}}}]{Li04}
\bibinfo{author}{\bibfnamefont{T.-P.} \bibnamefont{{Li}}},
  \bibinfo{author}{\bibfnamefont{J.-L.} \bibnamefont{{Qu}}},
  \bibinfo{author}{\bibfnamefont{H.}~\bibnamefont{{Feng}}},
  \bibinfo{author}{\bibfnamefont{L.-M.} \bibnamefont{{Song}}},
  \bibinfo{author}{\bibfnamefont{G.-Q.} \bibnamefont{{Ding}}},
  \bibnamefont{and} \bibinfo{author}{\bibfnamefont{L.}~\bibnamefont{{Chen}}},
  \bibinfo{journal}{\cjaa} \textbf{\bibinfo{volume}{4}}, \bibinfo{pages}{583}
  (\bibinfo{year}{2004}), \eprint{astro-ph/0407458}.

\bibitem[{\citenamefont{{Xiao} et~al.}(2021)\citenamefont{{Xiao}, {Xiong},
  {Zhang}, {Song}, {Lu}, {Huang}, {Cai}, {Yi}, {Song}, {Chen} et~al.}}]{Xiao21}
\bibinfo{author}{\bibfnamefont{S.}~\bibnamefont{{Xiao}}},
  \bibinfo{author}{\bibfnamefont{S.~L.} \bibnamefont{{Xiong}}},
  \bibinfo{author}{\bibfnamefont{S.~N.} \bibnamefont{{Zhang}}},
  \bibinfo{author}{\bibfnamefont{L.~M.} \bibnamefont{{Song}}},
  \bibinfo{author}{\bibfnamefont{F.~J.} \bibnamefont{{Lu}}},
  \bibinfo{author}{\bibfnamefont{Y.}~\bibnamefont{{Huang}}},
  \bibinfo{author}{\bibfnamefont{C.}~\bibnamefont{{Cai}}},
  \bibinfo{author}{\bibfnamefont{Q.~B.} \bibnamefont{{Yi}}},
  \bibinfo{author}{\bibfnamefont{X.~Y.} \bibnamefont{{Song}}},
  \bibinfo{author}{\bibfnamefont{W.}~\bibnamefont{{Chen}}},
  \bibnamefont{et~al.}, \bibinfo{journal}{\apj} \textbf{\bibinfo{volume}{920}},
  \bibinfo{eid}{43} (\bibinfo{year}{2021}).

\bibitem[{\citenamefont{{Krishak} and {Desai}}(2020)}]{Krishak4}
\bibinfo{author}{\bibfnamefont{A.}~\bibnamefont{{Krishak}}} \bibnamefont{and}
  \bibinfo{author}{\bibfnamefont{S.}~\bibnamefont{{Desai}}},
  \bibinfo{journal}{\jcap} \textbf{\bibinfo{volume}{2020}}, \bibinfo{eid}{006}
  (\bibinfo{year}{2020}), \eprint{2003.10127}.

\bibitem[{\citenamefont{{Wei} et~al.}(2017{\natexlab{a}})\citenamefont{{Wei},
  {Zhang}, {Shao}, {Wu}, and {M{\'e}sz{\'a}ros}}}]{Wei}
\bibinfo{author}{\bibfnamefont{J.-J.} \bibnamefont{{Wei}}},
  \bibinfo{author}{\bibfnamefont{B.-B.} \bibnamefont{{Zhang}}},
  \bibinfo{author}{\bibfnamefont{L.}~\bibnamefont{{Shao}}},
  \bibinfo{author}{\bibfnamefont{X.-F.} \bibnamefont{{Wu}}}, \bibnamefont{and}
  \bibinfo{author}{\bibfnamefont{P.}~\bibnamefont{{M{\'e}sz{\'a}ros}}},
  \bibinfo{journal}{\apjl} \textbf{\bibinfo{volume}{834}}, \bibinfo{eid}{L13}
  (\bibinfo{year}{2017}{\natexlab{a}}), \eprint{1612.09425}.

\bibitem[{\citenamefont{{Shao} et~al.}(2017)\citenamefont{{Shao}, {Zhang},
  {Wang}, {Wu}, {Cheng}, {Zhang}, {Yu}, {Xi}, {Wang}, {Feng} et~al.}}]{Shao}
\bibinfo{author}{\bibfnamefont{L.}~\bibnamefont{{Shao}}},
  \bibinfo{author}{\bibfnamefont{B.-B.} \bibnamefont{{Zhang}}},
  \bibinfo{author}{\bibfnamefont{F.-R.} \bibnamefont{{Wang}}},
  \bibinfo{author}{\bibfnamefont{X.-F.} \bibnamefont{{Wu}}},
  \bibinfo{author}{\bibfnamefont{Y.-H.} \bibnamefont{{Cheng}}},
  \bibinfo{author}{\bibfnamefont{X.}~\bibnamefont{{Zhang}}},
  \bibinfo{author}{\bibfnamefont{B.-Y.} \bibnamefont{{Yu}}},
  \bibinfo{author}{\bibfnamefont{B.-J.} \bibnamefont{{Xi}}},
  \bibinfo{author}{\bibfnamefont{X.}~\bibnamefont{{Wang}}},
  \bibinfo{author}{\bibfnamefont{H.-X.} \bibnamefont{{Feng}}},
  \bibnamefont{et~al.}, \bibinfo{journal}{\apj} \textbf{\bibinfo{volume}{844}},
  \bibinfo{eid}{126} (\bibinfo{year}{2017}), \eprint{1610.07191}.

\bibitem[{\citenamefont{{Perennes} et~al.}(2020)\citenamefont{{Perennes},
  {Sol}, and {Bolmont}}}]{Perennes20}
\bibinfo{author}{\bibfnamefont{C.}~\bibnamefont{{Perennes}}},
  \bibinfo{author}{\bibfnamefont{H.}~\bibnamefont{{Sol}}}, \bibnamefont{and}
  \bibinfo{author}{\bibfnamefont{J.}~\bibnamefont{{Bolmont}}},
  \bibinfo{journal}{\aap} \textbf{\bibinfo{volume}{633}}, \bibinfo{eid}{A143}
  (\bibinfo{year}{2020}), \eprint{1911.10377}.

\bibitem[{\citenamefont{Aghanim et~al.}(2020)}]{Planck18}
\bibinfo{author}{\bibfnamefont{N.}~\bibnamefont{Aghanim}} \bibnamefont{et~al.}
  (\bibinfo{collaboration}{Planck}), \bibinfo{journal}{Astron. Astrophys.}
  \textbf{\bibinfo{volume}{641}}, \bibinfo{pages}{A6} (\bibinfo{year}{2020}),
  \eprint{1807.06209}.

\bibitem[{\citenamefont{Abdalla et~al.}(2022)}]{Abdalla}
\bibinfo{author}{\bibfnamefont{E.}~\bibnamefont{Abdalla}} \bibnamefont{et~al.},
  \bibinfo{journal}{JHEAp} \textbf{\bibinfo{volume}{34}}, \bibinfo{pages}{49}
  (\bibinfo{year}{2022}), \eprint{2203.06142}.

\bibitem[{\citenamefont{{Trotta}}(2017)}]{Trotta}
\bibinfo{author}{\bibfnamefont{R.}~\bibnamefont{{Trotta}}},
  \bibinfo{journal}{ArXiv e-prints}  (\bibinfo{year}{2017}),
  \eprint{1701.01467}.

\bibitem[{\citenamefont{{Sharma}}(2017)}]{Sanjib}
\bibinfo{author}{\bibfnamefont{S.}~\bibnamefont{{Sharma}}},
  \bibinfo{journal}{\araa} \textbf{\bibinfo{volume}{55}}, \bibinfo{pages}{213}
  (\bibinfo{year}{2017}), \eprint{1706.01629}.

\bibitem[{\citenamefont{{Kerscher} and {Weller}}(2019)}]{Weller}
\bibinfo{author}{\bibfnamefont{M.}~\bibnamefont{{Kerscher}}} \bibnamefont{and}
  \bibinfo{author}{\bibfnamefont{J.}~\bibnamefont{{Weller}}},
  \bibinfo{journal}{SciPost Physics Lecture Notes} \textbf{\bibinfo{volume}{9}}
  (\bibinfo{year}{2019}), \eprint{1901.07726}.

\bibitem[{\citenamefont{{Bromberg} et~al.}(2013)\citenamefont{{Bromberg},
  {Nakar}, {Piran}, and {Sari}}}]{Bromberg}
\bibinfo{author}{\bibfnamefont{O.}~\bibnamefont{{Bromberg}}},
  \bibinfo{author}{\bibfnamefont{E.}~\bibnamefont{{Nakar}}},
  \bibinfo{author}{\bibfnamefont{T.}~\bibnamefont{{Piran}}}, \bibnamefont{and}
  \bibinfo{author}{\bibfnamefont{R.}~\bibnamefont{{Sari}}},
  \bibinfo{journal}{\apj} \textbf{\bibinfo{volume}{764}}, \bibinfo{eid}{179}
  (\bibinfo{year}{2013}), \eprint{1210.0068}.

\bibitem[{\citenamefont{{Evans} et~al.}(2009)\citenamefont{{Evans},
  {Beardmore}, {Page}, {Osborne}, {O'Brien}, {Willingale}, {Starling},
  {Burrows}, {Godet}, {Vetere} et~al.}}]{Evans09}
\bibinfo{author}{\bibfnamefont{P.~A.} \bibnamefont{{Evans}}},
  \bibinfo{author}{\bibfnamefont{A.~P.} \bibnamefont{{Beardmore}}},
  \bibinfo{author}{\bibfnamefont{K.~L.} \bibnamefont{{Page}}},
  \bibinfo{author}{\bibfnamefont{J.~P.} \bibnamefont{{Osborne}}},
  \bibinfo{author}{\bibfnamefont{P.~T.} \bibnamefont{{O'Brien}}},
  \bibinfo{author}{\bibfnamefont{R.}~\bibnamefont{{Willingale}}},
  \bibinfo{author}{\bibfnamefont{R.~L.~C.} \bibnamefont{{Starling}}},
  \bibinfo{author}{\bibfnamefont{D.~N.} \bibnamefont{{Burrows}}},
  \bibinfo{author}{\bibfnamefont{O.}~\bibnamefont{{Godet}}},
  \bibinfo{author}{\bibfnamefont{L.}~\bibnamefont{{Vetere}}},
  \bibnamefont{et~al.}, \bibinfo{journal}{\mnras}
  \textbf{\bibinfo{volume}{397}}, \bibinfo{pages}{1177} (\bibinfo{year}{2009}),
  \eprint{0812.3662}.

\bibitem[{\citenamefont{{von Kienlin} et~al.}(2020)\citenamefont{{von Kienlin},
  {Meegan}, {Paciesas}, {Bhat}, {Bissaldi}, {Briggs}, {Burns}, {Cleveland},
  {Gibby}, {Giles} et~al.}}]{VonKienlin}
\bibinfo{author}{\bibfnamefont{A.}~\bibnamefont{{von Kienlin}}},
  \bibinfo{author}{\bibfnamefont{C.~A.} \bibnamefont{{Meegan}}},
  \bibinfo{author}{\bibfnamefont{W.~S.} \bibnamefont{{Paciesas}}},
  \bibinfo{author}{\bibfnamefont{P.~N.} \bibnamefont{{Bhat}}},
  \bibinfo{author}{\bibfnamefont{E.}~\bibnamefont{{Bissaldi}}},
  \bibinfo{author}{\bibfnamefont{M.~S.} \bibnamefont{{Briggs}}},
  \bibinfo{author}{\bibfnamefont{E.}~\bibnamefont{{Burns}}},
  \bibinfo{author}{\bibfnamefont{W.~H.} \bibnamefont{{Cleveland}}},
  \bibinfo{author}{\bibfnamefont{M.~H.} \bibnamefont{{Gibby}}},
  \bibinfo{author}{\bibfnamefont{M.~M.} \bibnamefont{{Giles}}},
  \bibnamefont{et~al.}, \bibinfo{journal}{\apj} \textbf{\bibinfo{volume}{893}},
  \bibinfo{eid}{46} (\bibinfo{year}{2020}), \eprint{2002.11460}.

\bibitem[{\citenamefont{{Band} et~al.}(1993)\citenamefont{{Band}, {Matteson},
  {Ford}, {Schaefer}, {Palmer}, {Teegarden}, {Cline}, {Briggs}, {Paciesas},
  {Pendleton} et~al.}}]{Band}
\bibinfo{author}{\bibfnamefont{D.}~\bibnamefont{{Band}}},
  \bibinfo{author}{\bibfnamefont{J.}~\bibnamefont{{Matteson}}},
  \bibinfo{author}{\bibfnamefont{L.}~\bibnamefont{{Ford}}},
  \bibinfo{author}{\bibfnamefont{B.}~\bibnamefont{{Schaefer}}},
  \bibinfo{author}{\bibfnamefont{D.}~\bibnamefont{{Palmer}}},
  \bibinfo{author}{\bibfnamefont{B.}~\bibnamefont{{Teegarden}}},
  \bibinfo{author}{\bibfnamefont{T.}~\bibnamefont{{Cline}}},
  \bibinfo{author}{\bibfnamefont{M.}~\bibnamefont{{Briggs}}},
  \bibinfo{author}{\bibfnamefont{W.}~\bibnamefont{{Paciesas}}},
  \bibinfo{author}{\bibfnamefont{G.}~\bibnamefont{{Pendleton}}},
  \bibnamefont{et~al.}, \bibinfo{journal}{\apj} \textbf{\bibinfo{volume}{413}},
  \bibinfo{pages}{281} (\bibinfo{year}{1993}).

\bibitem[{\citenamefont{{Wei} et~al.}(2017{\natexlab{b}})\citenamefont{{Wei},
  {Wu}, {Zhang}, {Shao}, {M{\'e}sz{\'a}ros}, and {Kosteleck{\'y}}}}]{Wei2}
\bibinfo{author}{\bibfnamefont{J.-J.} \bibnamefont{{Wei}}},
  \bibinfo{author}{\bibfnamefont{X.-F.} \bibnamefont{{Wu}}},
  \bibinfo{author}{\bibfnamefont{B.-B.} \bibnamefont{{Zhang}}},
  \bibinfo{author}{\bibfnamefont{L.}~\bibnamefont{{Shao}}},
  \bibinfo{author}{\bibfnamefont{P.}~\bibnamefont{{M{\'e}sz{\'a}ros}}},
  \bibnamefont{and} \bibinfo{author}{\bibfnamefont{V.~A.}
  \bibnamefont{{Kosteleck{\'y}}}}, \bibinfo{journal}{\apj}
  \textbf{\bibinfo{volume}{842}}, \bibinfo{eid}{115}
  (\bibinfo{year}{2017}{\natexlab{b}}), \eprint{1704.05984}.

\bibitem[{\citenamefont{{Fishman}}(1995)}]{Fishman95}
\bibinfo{author}{\bibfnamefont{G.~J.} \bibnamefont{{Fishman}}},
  \bibinfo{journal}{\pasp} \textbf{\bibinfo{volume}{107}},
  \bibinfo{pages}{1145} (\bibinfo{year}{1995}).

\bibitem[{\citenamefont{{Vardanyan} et~al.}(2022)\citenamefont{{Vardanyan},
  {Takhistov}, {Ata}, and {Murase}}}]{Murase22}
\bibinfo{author}{\bibfnamefont{V.}~\bibnamefont{{Vardanyan}}},
  \bibinfo{author}{\bibfnamefont{V.}~\bibnamefont{{Takhistov}}},
  \bibinfo{author}{\bibfnamefont{M.}~\bibnamefont{{Ata}}}, \bibnamefont{and}
  \bibinfo{author}{\bibfnamefont{K.}~\bibnamefont{{Murase}}},
  \bibinfo{journal}{arXiv e-prints} \bibinfo{eid}{arXiv:2212.02436}
  (\bibinfo{year}{2022}), \eprint{2212.02436}.

\bibitem[{\citenamefont{{Jacob} and {Piran}}(2008)}]{Jacob}
\bibinfo{author}{\bibfnamefont{U.}~\bibnamefont{{Jacob}}} \bibnamefont{and}
  \bibinfo{author}{\bibfnamefont{T.}~\bibnamefont{{Piran}}},
  \bibinfo{journal}{\jcap} \textbf{\bibinfo{volume}{1}}, \bibinfo{eid}{031}
  (\bibinfo{year}{2008}), \eprint{0712.2170}.

\bibitem[{\citenamefont{Seikel et~al.}(2012)\citenamefont{Seikel, Clarkson, and
  Smith}}]{Seikel_2012}
\bibinfo{author}{\bibfnamefont{M.}~\bibnamefont{Seikel}},
  \bibinfo{author}{\bibfnamefont{C.}~\bibnamefont{Clarkson}}, \bibnamefont{and}
  \bibinfo{author}{\bibfnamefont{M.}~\bibnamefont{Smith}},
  \bibinfo{journal}{JCAP} \textbf{\bibinfo{volume}{1206}}, \bibinfo{pages}{036}
  (\bibinfo{year}{2012}), \eprint{1204.2832}.

\bibitem[{\citenamefont{{Jimenez} and {Loeb}}(2002)}]{Jimenez_2002}
\bibinfo{author}{\bibfnamefont{R.}~\bibnamefont{{Jimenez}}} \bibnamefont{and}
  \bibinfo{author}{\bibfnamefont{A.}~\bibnamefont{{Loeb}}},
  \bibinfo{journal}{\apj} \textbf{\bibinfo{volume}{573}}, \bibinfo{pages}{37}
  (\bibinfo{year}{2002}), \eprint{astro-ph/0106145}.

\bibitem[{\citenamefont{{Singirikonda} and {Desai}}(2020)}]{Haveesh}
\bibinfo{author}{\bibfnamefont{H.}~\bibnamefont{{Singirikonda}}}
  \bibnamefont{and} \bibinfo{author}{\bibfnamefont{S.}~\bibnamefont{{Desai}}},
  \bibinfo{journal}{European Physical Journal C} \textbf{\bibinfo{volume}{80}},
  \bibinfo{eid}{694} (\bibinfo{year}{2020}), \eprint{2003.00494}.

\bibitem[{\citenamefont{{Bora} and {Desai}}(2021)}]{BoraCDDR}
\bibinfo{author}{\bibfnamefont{K.}~\bibnamefont{{Bora}}} \bibnamefont{and}
  \bibinfo{author}{\bibfnamefont{S.}~\bibnamefont{{Desai}}},
  \bibinfo{journal}{\jcap} \textbf{\bibinfo{volume}{2021}}, \bibinfo{eid}{052}
  (\bibinfo{year}{2021}), \eprint{2104.00974}.

\bibitem[{\citenamefont{{Vagnozzi} et~al.}(2021)\citenamefont{{Vagnozzi},
  {Loeb}, and {Moresco}}}]{Vagnozzi}
\bibinfo{author}{\bibfnamefont{S.}~\bibnamefont{{Vagnozzi}}},
  \bibinfo{author}{\bibfnamefont{A.}~\bibnamefont{{Loeb}}}, \bibnamefont{and}
  \bibinfo{author}{\bibfnamefont{M.}~\bibnamefont{{Moresco}}},
  \bibinfo{journal}{\apj} \textbf{\bibinfo{volume}{908}}, \bibinfo{eid}{84}
  (\bibinfo{year}{2021}), \eprint{2011.11645}.

\bibitem[{\citenamefont{{Holanda} et~al.}(2022)\citenamefont{{Holanda}, {Bora},
  and {Desai}}}]{Bora22}
\bibinfo{author}{\bibfnamefont{R.~F.~L.} \bibnamefont{{Holanda}}},
  \bibinfo{author}{\bibfnamefont{K.}~\bibnamefont{{Bora}}}, \bibnamefont{and}
  \bibinfo{author}{\bibfnamefont{S.}~\bibnamefont{{Desai}}},
  \bibinfo{journal}{European Physical Journal C} \textbf{\bibinfo{volume}{82}},
  \bibinfo{eid}{526} (\bibinfo{year}{2022}), \eprint{2105.10988}.

\bibitem[{\citenamefont{{Moresco} et~al.}(2022)\citenamefont{{Moresco},
  {Amati}, {Amendola}, {Birrer}, {Blakeslee}, {Cantiello}, {Cimatti},
  {Darling}, {Della Valle}, {Fishbach} et~al.}}]{Moresco}
\bibinfo{author}{\bibfnamefont{M.}~\bibnamefont{{Moresco}}},
  \bibinfo{author}{\bibfnamefont{L.}~\bibnamefont{{Amati}}},
  \bibinfo{author}{\bibfnamefont{L.}~\bibnamefont{{Amendola}}},
  \bibinfo{author}{\bibfnamefont{S.}~\bibnamefont{{Birrer}}},
  \bibinfo{author}{\bibfnamefont{J.~P.} \bibnamefont{{Blakeslee}}},
  \bibinfo{author}{\bibfnamefont{M.}~\bibnamefont{{Cantiello}}},
  \bibinfo{author}{\bibfnamefont{A.}~\bibnamefont{{Cimatti}}},
  \bibinfo{author}{\bibfnamefont{J.}~\bibnamefont{{Darling}}},
  \bibinfo{author}{\bibfnamefont{M.}~\bibnamefont{{Della Valle}}},
  \bibinfo{author}{\bibfnamefont{M.}~\bibnamefont{{Fishbach}}},
  \bibnamefont{et~al.}, \bibinfo{journal}{arXiv e-prints}
  \bibinfo{eid}{arXiv:2201.07241} (\bibinfo{year}{2022}), \eprint{2201.07241}.

\bibitem[{\citenamefont{{Rosati} et~al.}(2015)\citenamefont{{Rosati},
  {Amelino-Camelia}, {Marcian{\`o}}, and {Matassa}}}]{Rosati}
\bibinfo{author}{\bibfnamefont{G.}~\bibnamefont{{Rosati}}},
  \bibinfo{author}{\bibfnamefont{G.}~\bibnamefont{{Amelino-Camelia}}},
  \bibinfo{author}{\bibfnamefont{A.}~\bibnamefont{{Marcian{\`o}}}},
  \bibnamefont{and}
  \bibinfo{author}{\bibfnamefont{M.}~\bibnamefont{{Matassa}}},
  \bibinfo{journal}{\prd} \textbf{\bibinfo{volume}{92}}, \bibinfo{eid}{124042}
  (\bibinfo{year}{2015}), \eprint{1507.02056}.

\bibitem[{\citenamefont{{Bolmont} et~al.}(2022)\citenamefont{{Bolmont},
  {Caroff}, {Gaug}, {Gent}, {Jacholkowska}, {Kerszberg}, {Levy}, {Lin},
  {Martinez}, {Nogu{\'e}s} et~al.}}]{Levy21}
\bibinfo{author}{\bibfnamefont{J.}~\bibnamefont{{Bolmont}}},
  \bibinfo{author}{\bibfnamefont{S.}~\bibnamefont{{Caroff}}},
  \bibinfo{author}{\bibfnamefont{M.}~\bibnamefont{{Gaug}}},
  \bibinfo{author}{\bibfnamefont{A.}~\bibnamefont{{Gent}}},
  \bibinfo{author}{\bibfnamefont{A.}~\bibnamefont{{Jacholkowska}}},
  \bibinfo{author}{\bibfnamefont{D.}~\bibnamefont{{Kerszberg}}},
  \bibinfo{author}{\bibfnamefont{C.}~\bibnamefont{{Levy}}},
  \bibinfo{author}{\bibfnamefont{T.}~\bibnamefont{{Lin}}},
  \bibinfo{author}{\bibfnamefont{M.}~\bibnamefont{{Martinez}}},
  \bibinfo{author}{\bibfnamefont{L.}~\bibnamefont{{Nogu{\'e}s}}},
  \bibnamefont{et~al.}, \bibinfo{journal}{\apj} \textbf{\bibinfo{volume}{930}},
  \bibinfo{eid}{75} (\bibinfo{year}{2022}), \eprint{2201.02087}.

\bibitem[{\citenamefont{{Krishak} et~al.}(2020)\citenamefont{{Krishak},
  {Dantuluri}, and {Desai}}}]{Krishak1}
\bibinfo{author}{\bibfnamefont{A.}~\bibnamefont{{Krishak}}},
  \bibinfo{author}{\bibfnamefont{A.}~\bibnamefont{{Dantuluri}}},
  \bibnamefont{and} \bibinfo{author}{\bibfnamefont{S.}~\bibnamefont{{Desai}}},
  \bibinfo{journal}{\jcap} \textbf{\bibinfo{volume}{2020}}, \bibinfo{eid}{007}
  (\bibinfo{year}{2020}), \eprint{1906.05726}.

\bibitem[{\citenamefont{{Bhagvati} and {Desai}}(2021)}]{Srinikitha}
\bibinfo{author}{\bibfnamefont{S.}~\bibnamefont{{Bhagvati}}} \bibnamefont{and}
  \bibinfo{author}{\bibfnamefont{S.}~\bibnamefont{{Desai}}},
  \bibinfo{journal}{\jcap} \textbf{\bibinfo{volume}{2021}}, \bibinfo{eid}{022}
  (\bibinfo{year}{2021}), \eprint{2106.06724}.

\bibitem[{\citenamefont{{Speagle}}(2020)}]{dynesty}
\bibinfo{author}{\bibfnamefont{J.~S.} \bibnamefont{{Speagle}}},
  \bibinfo{journal}{\mnras}  (\bibinfo{year}{2020}), \eprint{1904.02180}.

\bibitem[{\citenamefont{Group et~al.}(2020)\citenamefont{Group, Zyla, Barnett,
  Beringer, Dahl, Dwyer, Groom, Lin, Lugovsky, Pianori et~al.}}]{PDG}
\bibinfo{author}{\bibfnamefont{P.~D.} \bibnamefont{Group}},
  \bibinfo{author}{\bibfnamefont{P.}~\bibnamefont{Zyla}},
  \bibinfo{author}{\bibfnamefont{R.}~\bibnamefont{Barnett}},
  \bibinfo{author}{\bibfnamefont{J.}~\bibnamefont{Beringer}},
  \bibinfo{author}{\bibfnamefont{O.}~\bibnamefont{Dahl}},
  \bibinfo{author}{\bibfnamefont{D.}~\bibnamefont{Dwyer}},
  \bibinfo{author}{\bibfnamefont{D.}~\bibnamefont{Groom}},
  \bibinfo{author}{\bibfnamefont{C.-J.} \bibnamefont{Lin}},
  \bibinfo{author}{\bibfnamefont{K.}~\bibnamefont{Lugovsky}},
  \bibinfo{author}{\bibfnamefont{E.}~\bibnamefont{Pianori}},
  \bibnamefont{et~al.}, \bibinfo{journal}{Progress of Theoretical and
  Experimental Physics} \textbf{\bibinfo{volume}{2020}},
  \bibinfo{pages}{083C01} (\bibinfo{year}{2020}).

\bibitem[{\citenamefont{{Biesiada} and {Pi{\'o}rkowska}}(2009)}]{Marek10}
\bibinfo{author}{\bibfnamefont{M.}~\bibnamefont{{Biesiada}}} \bibnamefont{and}
  \bibinfo{author}{\bibfnamefont{A.}~\bibnamefont{{Pi{\'o}rkowska}}},
  \bibinfo{journal}{Classical and Quantum Gravity}
  \textbf{\bibinfo{volume}{26}}, \bibinfo{eid}{125007} (\bibinfo{year}{2009}),
  \eprint{1008.2615}.

\bibitem[{\citenamefont{{Chang} et~al.}(2012)\citenamefont{{Chang}, {Jiang},
  and {Lin}}}]{Chang12}
\bibinfo{author}{\bibfnamefont{Z.}~\bibnamefont{{Chang}}},
  \bibinfo{author}{\bibfnamefont{Y.}~\bibnamefont{{Jiang}}}, \bibnamefont{and}
  \bibinfo{author}{\bibfnamefont{H.-N.} \bibnamefont{{Lin}}},
  \bibinfo{journal}{Astroparticle Physics} \textbf{\bibinfo{volume}{36}},
  \bibinfo{pages}{47} (\bibinfo{year}{2012}), \eprint{1201.3413}.

\end{thebibliography}
\end{document}